\documentclass[manuscript]{aastex}

\slugcomment{To appear in the Astronomical Journal}
\shorttitle{Near-IR NICI Images of Eta Car}
\shortauthors{Artigau et al.}  

\begin{document}


\title{Penetrating the Homunculus -- Near-Infrared Adaptive Optics Images of Eta Carinae 
\altaffilmark{1}}


\author{\'Etienne Artigau}
\affil{Gemini Observatory - South and D\'epartement de Physique and Observatoire du Mont M\'egantic, Universit\'e de Montr\'eal, QC, H3C 3J7, Canada}
\author{John C. Martin}
\affil{Department of Physics and Astronomy, University of Illinois -- Springfield, 62703}
\author{Roberta M. Humphreys}
\author{Kris Davidson}
\affil{Astronomy Department, University of Minnesota,
   55455}
\author{Olivier Chesneau}
\affil{Observatoire de la C\^ote d'Azur, France}
\and 
\author{Nathan Smith\altaffilmark{2}}
\affil{Astronomy Department, University of California, Berkeley, 94720}

\altaffiltext{1}{Based on observations obtained at the Gemini Observatory (program ID : GS-2008B-DD-6),
which is operated by the Association of Universities for Research in Astronomy, Inc.,
under a cooperative agreement with the NSF on behalf of the Gemini partnership: the
National Science Foundation (United States), the Science and Technology Facilities
Council (United Kingdom), the National Research Council (Canada), CONICYT (Chile), the
Australian Research Council (Australia), Minist\'{e}rio da Ci\^{e}ncia e Tecnologia
(Brazil) and Ministerio de Ciencia, Tecnolog\'{i}a e Innovaci\'{o}n Productiva
(Argentina). } 
\altaffiltext{2}{Now at Steward Observatory, University of Arizona, Tucson, 85721}

\begin{abstract}
Near-infrared adaptive optics imaging with NICI and NaCO reveal what appears to be a
three-winged or lobed pattern, the ``butterfly nebula'', outlined by bright Br$\gamma$ and H$_{2}$ emission and light scattered by 
dust. In contrast, the [Fe II] emission does not follow the 
outline of the wings, but shows an extended bipolar distribution which is tracing the 
Little Homunculus ejected in $\eta$ Car's second or lesser eruption in the 1890's.
Proper motions measured from the combined NICI and NaCO images together with radial
velocities show that the knots and filaments that define the bright rims of the butterfly
were ejected at two different epochs corresponding approximately to  the great eruption 
and the second eruption. Most of the material is spatially distributed 10$\arcdeg$ to 20$\arcdeg$ 
above and below 
the equatorial plane apparently behind the Little Homunculus and 
the larger SE lobe. The equatorial debris either has a wide opening angle or the 
clumps were ejected at different latitudes relative to the plane. The butterfly is
not a coherent physical structure or equatorial torus but spatially separate clumps and
filaments ejected at different times, and now 2000 to 4000 AU from the star. 
\end{abstract}


\keywords{circumstellar matter -- ISM:individual(Homunculus Nebula) -- ISM:jets and outflows -- stars:individual($\eta$ Carinae) -- stars:mass loss -- stars:winds, outflows}

\section{Introduction}

Eta Carinae's dusty, mottled, bipolar Homunculus nebula and its ragged equatorial ``skirt'' of debris provide a visual record of its two eruptions in the 19th century. Numerous studies to measure the proper motions and expansion of the ejecta have shown that the bipolar lobes were ejected during $\eta$ Car's ``great eruption'' in the 1840's (Gaviola 1950, Ringuelet 1958, Currie et al. 1996, Smith \& Gehrz 1998, Morse et al. 2001).  Proper motion and radial velocity measurements of the equatorial ejecta revealed material very likely expelled in two separate eruptions \citep{KD97,KD01,Smith98,Currie99,Morse01}, in the 1840's and in the lesser known second eruption in the 1890's (see Humphreys, Davidson and Smith 1999). A second, smaller bipolar outflow, now called the ``Little Homunculus'', was discovered by Ishibashi et al. (2003) based on the Doppler--shifted morphology of the integral-sign shaped Fe II and [Fe II] emission lines in  HST/STIS spectra. It is embedded in the Homunculus  and extends $\sim 2\arcsec$ in either direction along the major axis. Its kinematics suggest an ejection date associated with the 1890's eruption \citep{Bish03,Bish05,Smith05}. The Little Homunculus cannot be distinguished in  visual images of $\eta$ Car due to bright nebulosity near the star and to  extinction in the Homunculus, but its structure can be traced by the velocities of the  [Fe II] emission lines (Ishibashi et al. 2003; Smith 2002, 2005; Teodoro et al. 2008).

Infrared imaging and spectroscopy allow us to penetrate the dusty outer shell of the Homunculus to reveal a complex structure near the central star. Mid-infrared imaging, from 5$\mu$m to 24$\mu$m, revealed an elongated feature  several arcseconds across centered on the star and interpreted as a circumstellar ring (Rigaut \& Gehring 1995; Smith et al. 1998; Polomski et al. 1999; Smith \& Gehrz 2000). The highest resolution mid-IR imaging to date, \citep{CSmith95,Smith02b,Ches05},  show what looks like two interlocking rings or loops with multiple bright knots. Near-IR images (1 -- 2 $\mu$m) obtained  with HST/NICMOS were discussed by  Smith and Gehrz (1998) who identified a loop-like structure which  they also attributed to an equatorial torus. A  near-IR (1 -- 4$\mu$m) study by  Chesneau et al (2005) using the VLT adaptive optics system, NaCO\footnote{The Nasmyth Adaptive Optics System and Near-Infrared Imager and
 Spectrograph on UT4 of the VLT under programs 074.D-0140, 076.D-0586, 078.D-0562.}
resolved this feature into  a complex  dusty structure, the ``butterfly'' nebula, with three fans or wings  outlined by emitting warm dust and reflected hydrogen emission, around the relatively dark wings which are apparently devoid of dust. The wings of the butterfly appear to be emanating from the region of  the central star. Chesneau et al. suggested that this was a dusty, bipolar nebula distinct from the little Homunculus. Alternatively,  Smith (2006)  has argued for a distorted equatorial torus that has been swept back by the stellar wind. Kinematic data are needed to determine the geometric orientation of this complex structure with respect to the axis of symmetry of the Homunculus and the equatorial ejecta and its ejection epoch. 

We obtained narrow-band images with the Near-Infrared Coronagraphic Imager (NICI) and adaptive-optics on the Gemini-South telescope just after $\eta$ Car's 2009 spectroscopic event. In the next section we describe the observing procedure and data reduction. The narrow-band  Br$\gamma$, H$_{2}$, and [FeII] images are discussed in \S {3}. The Br$\gamma$ and H$_{2}$ images are essentially identical. Both trace the outlines of the butterfly nebula, while the continuum-subtracted [Fe II] image 
reveals a bipolar distribution due to the little Homunculus. In \S {4} we compare our images with the 
NaCO images obtained several years earlier and measure the  projected motions. The expansion, orientation, and ejection epoch for the resolved features in the butterfly nebula are discussed in the last section.

\section{Observations and Data Reduction} 

We used the Near Infrared Coronagraphic Imager (NICI) (Ftaclas, Mart\'in \& Toomey, 2003, Chun et al. 2008) to image the Homunculus of $\eta$ Car on 2009 February, 11 UT, a few weeks after the spectroscopic minimum. 
NICI features a dual channel camera, primarily designed for Spectral Differential Imaging (SDI) for planet searches (Marois et al., 2005, Biller et al., 2004), 
but the simultaneity of the two channel imaging also provides optimal {\it on/off} line imaging. This simultaneity ensures that the observing conditions (transmission, seeing) for the two channels are matched. Because of its extended nebulosity, the SDI technique was not appropriate for $\eta$ Car.  

Pairs of exposures were recorded simultaneously in the two channels with different filters. The Cassegrain rotator was turned off during the observations. The filters were chosen to map the spatial extent of previously identified emission features in $\eta$ Car's near-infrared spectrum and to provide {\it off}-band continuum images for subtraction. Near-infrared spectra \citep{Smith01} were used to determine which filters would yield {\it off}-band continuum images free of strong emission features. The filters  
and their characteristics are listed in Table 1.  The exposure in each filter is a coaddition of 
thirty $0\farcs38$\,s subexposures.
During our observations the seeing varied between $0\farcs6$ and $1\farcs0$. The effective resolution was $0\farcs056$ and $0\farcs060$ for the $H$ and $K$-band filters, respectively, close to the corresponding diffraction limits of $0\farcs044$ and $0\farcs057$. NICI has an $0\farcs018$ pixel scale providing a Nyquist sampling of the $H$ and $K$-band diffraction limits. We used the 95\% pupil mask with the $0\farcs32$ semi-transparent focal plane mask to block the central star. The mask slightly degrades the resolution, but minimizes the intensity of the speckle pattern. 
With the rotator turned off, the field slowly rotated  (6-8$^\circ$) through the observing sequence. The field rotation was used for spatial dithering and for bad pixel removal. This approach is more efficient than a classical dithering pattern as it avoids the overheads involved in the reacquisition behind the coronagraphic mask

Data reduction steps followed a relatively classical approach. The images were first dark-subtracted and flat-fielded using a median-combined flat that was produced using flat frames taken without the coronagraphic mask. The images were then registered using the unsaturated core of the central star, visible through the semi-transparent coronagraphic mask. A ``bad pixel'' mask was produced from the flat field images, and bad pixels were flagged in all images. Images were then derotated to a common position angle using the known parallactic angle at the time of observation. The images were median combined to produce the final image. This procedure was repeated for the {\it on} and {\it off} band filters and the resulting images were subtracted to produce a $line$ image. Figure 1 shows the resulting continuum and continuum-subtracted line images for each of the three filter combinations. The combined color continuum and line-continuum images are shown in Figures 2a and 2b. 

To better characterize small scale structures in the images, a set of final images
was produced by subtracting an $11\times11$ median filter from  all of the images before the final median combination. These high-passed images better highlight the filamentary structures in the inner Homunculus and are used later to measure their motions.

 The near-infrared NaCO images used in this paper were obtained between 2002 and 2006 by Chesneau and his collaborators. A new detector (InSb Aladdin 3 array) was installed in May 2004, replacing the Aladdin 2 detector that had 
been used since the camera was first available (Chesneau et al. 2005).
The NB\_374, NB\_405 narrow band filters\footnote{Central wavelengths 3.74$\mu$m and 4.05$\mu$m and 0.02$\mu$m wide.} were used to achieve the best 
optimization and contrast for the dusty neighborhood of $\eta$ Car.  
 The field of view was $28\arcsec\times28\arcsec$ 
and the scale was 27.1~mas per pixel. The reading mode was the so-called 
'Uncorr' mode in which the array is reset and then read once. The minimum 
effective integration time  
used for these observations was 0.1750\,s. The AutoJitter mode was used 
so that  at each exposure, the telescope moves according to a random pattern in a 
10$\arcsec$ box.  
The images were reduced with a self-developed IDL routine that processes the 
individual frames as described in Chesneau et al. 2005.

\section{Description of the Images -- Penetrating the Homunculus Nebula}

Our $K$-band continuum image and continuum subtracted narrow band Br$\gamma$ and H$_{2}$ images in Figure 1 and the color continuum image in Figure 2a 
clearly show the three winged dark pattern  named the butterfly nebula by
Chesneau et al. (2005). The  dark wings in both near and mid-infrared images are attributed to the lack of emitting dust at
these locations. The continuum-subtracted narrow-band Br$\gamma$ and H$_{2}$ images (Fig.1) show essentially identical patterns of emission due to scattered light by dust.  In Figure 3, we show our NICI $K$-band continuum image together with the NICMOS narrow-band 2.15$\mu$m image, at lower resolution, from Smith and Gehrz (2000) and the central region of the butterfly from the NaCO and NICI $K$-band images. Interestingly,  comparison of the two high resolution $K$-band images show no apparent change in the morphology  of the butterfly or of the warm infrared emitting regions.

In contrast with the Br$\gamma$ and H$_{2}$ images, the [Fe II] emission shows a very different distribution in Figure 1 and in the combined line color image in Figure 2a.
Although the corresponding continuum image shows a weak  butterfly pattern, the [Fe II] emission  does not follow the outline of the wings, but instead shows a much broader bipolar distribution that is most easily seen in Figure 2b. The spatial extent of the [Fe II] emission is readily traced in the nearer SE lobe out to $\sim$ 2$\arcsec$ from the star. We also note that a mottled structure is detectable in the SE lobe [Fe II] emission probably due to reflection in the overlying dusty Homunculus. A relatively bright emission region just above the mask is probably from the Weigelt knots where the [Fe II] emission is known to be strong (Davidson et al. 1995, 1997, Smith 2002).  Directly to the NW of the star we are very likely detecting a contribution to the [Fe II] emission from the ``fan''. Despite its appearance in the visual images of the Homunculus, this feature is not part of the equatorial spray, but instead a region where we see light from the star's north polar region directly reflected off the dust. Although fainter in the NW lobe, the  [Fe II] emission can also be traced to about 2$\arcsec$  along the major axis.

As we noted in the Introduction, the physical extent and shape of the 
little Homunculus is traced spectroscopically by its Fe II and [Fe II] emission. Figure 8 in 
the discovery paper \citep{Bish03} illustrates the  shape and spatial extent of the bipolar [Fe II] emission from the Little Homunculus relative to the Homunculus lobes. A velocity map of the Homunculus  \citep{Smith06} in the 1.64$\mu$m [Fe II] line, the same line imaged with NICI, and the IFU map by Teodoro et al. (2008) likewise
show the extent of the Little Homunculus to 2$\arcsec$ along the major axis.  The spectroscopic map also exihibits the same asymmetry in its brightness between the SE and NW lobes as mentioned above. We therefore conclude that the [Fe II] emission in the NICI image is tracing the Little Homunculus.

One of our original goals for high-resolution images near the time of $\eta$ Car's 2009 spectroscopic event was to look for possible changes in its dusty environment that may be due to changing illumination of the inner ejecta and excitation of the Weigelt knots as the secondary star passed through the massive primary's wind.  The Weigelt knots are small blobs or condensations of gas and dust $\approx$ 700 AU (0$\farcs$3) to the northwest of the star first identified by speckle 
interferometry \citep{Weig86}. The high excitation emission lines seen in ground-based spectra originate in the 
knots presumably due to UV radiation from the hot secondary star \citep{Mehner}. During the  spectroscopic events these emission lines weaken and disappear. 

As already noted there is no obvious variation in the $K$-band images obtained nearly seven years 
apart.  Due to the focal plane mask, the region of the Weigelt knots, where we might expect to observe some changes, is unfortunately not visible in the NICI images although the extremely IR-bright central star can be seen through the mask in the continuum images.  Small changes in positions of the 
knots had been detected in  speckle interferometry 
\citep{Weig95,Weig96} and later in HST imaging and spectroscopy \citep{KD97,Dor04,Smith04}.
The knots are moving very slowly, compared with the expansion of the Homunculus,
at  less than 50 km s$^{-1}$ (0$\farcs$005 yr$^{-1}$) in the equatorial plane. 
Fortunately no mask was used for the five  3.74$\mu$m (Pf$\gamma$) and 4.05$\mu$m (Br$\alpha$) NaCO 
images obtained from 2002 to 2006 and which bracket the 2003.5 event.  
The structures seen in the near-infrared images closely correspond with the knots  
seen in the visible or UV although they are not spatially coincident. The visible structures, 
dominated by scattering, trace the  {\it walls} of the dense clumps of dust, while the 
infrared structures are identified with the  emission from hot dust, probably the external 
layers of the clumps. The infrared condensations 
  corresponding to knots C and D, are therefore called  C$^{'}$ and D$^{'}$, see Chesnaeu et al. (2005). 

We carefully measured  the photocenter of the two most prominent knots, C$^{'}$  and D$^{'}$ in the 3.74$\mu$m images at position angles 300$\arcdeg$ and 352$\arcdeg$, respectively,
although the extraction of the centers was hampered by the increasing brightness
 of the star \citep{JCM} and Strehl ratio variations. The cuts through the clumps showing the 
  variation of intensity vs distance at different times are in 
 Figure 4. The normalized infrared flux appears to be decreasing from 2002 to 2005, although
 we emphasize that accurate photometry is difficult with the adaptive optics system. The decrease does not appear to be correlated with the 2003.5 event or
 orbital phase.  One possibility for the apparent decrease in the IR flux is enhanced dust destruction in response to the increased stellar flux, although this will have to be verified.

The interpretation of the complex  structure of the butterfly, with its large asymmetrical regions 
outlined by IR bright emission embedded in the Homunculus, is not straightforward. Chesneau et al. (2005) initially suggested that the emission rims of the butterfly
 wings are representative of an outflow along the polar axis projected onto the plane of the sky. Their 8.7$\mu$m image shows bright rims and  ``dark'' regions that correspond approximately with the  wings in the NaCO images which they ascribe to 
 minimal dust emission. Smith (2006) however demonstrated that the H$_{2}$ spectroscopic emission that outlines the thin walls of the Homunculus does not penetrate to the star but appears to terminate at the IR bright edges of the dark structure seen in his 8.8$\mu$m image. He therefore proposed that this structure is actually a disrupted equatorial torus in which the current equatorial stellar wind is clearing the equatorial plane of dust. Its complicated shape was then attributed to interaction with dusty clumps of different densities.

To further understand this peculiar structure, kinematic data is needed to measure or set limits on its expansion, age, and orientation within the Homunculus. In the next section we combine our NICI images with the comparable NaCO images obtained six to seven years earlier, to measure the projected motions of the knots and filaments that outline the butterfly.

\section{Measurement of the Projected Motions}

We registered  all of the NaCO images (K$_{s}$, 4.05 and 3.74$\mu$m) to the NICI
K-band continuum image by first rotating the images to a common orientation and performing a first-order scaling based on 
the known relative plate scales of NICI and NaCO. The initial, relative positioning was done
using the unsaturated core of the central star's psf. The precise scale and rotation were then 
fine-tuned iteratively  by maximizing the
correlation of the high-pass images in an annular region between $2\farcs3$ and 
$3\farcs5$ from the star.
This region contains structures that are visible in all of the  NaCO and NICI images, 
but is
sufficiently far from the butterfly to avoid the inner regions where we want to 
look for anomalous  motions. Figure 5 shows the scale vs the date of the 
NaCO and NICI images. 
Using the scaling factors and  the known pixel scales of NACO and NICI,
we then  derived the expansion rate and time since the eruption as a check on our 
procedure.  
Using data in the range $2\farcs3$ to $3\farcs5$ in all of the NaCO and NICI images, we find that  the Homunculus  lobes 
were ejected in  $1850.0\pm4.0$. Using different image combinations, NaCO 3.74$\mu$m 
with NICI, NaCO 4.05$\mu$m with NICI, and the NaCO data only, we obtain  similar results
within 
$1\sigma$ of the above value. This result is consistent with $\eta$ Car's
great eruption (1838--1858) and previous determinations of the expansion of the lobes
cited in \S {1}. 

Using the registered high-pass  NICI and the 2002.9 NaCO Ks image, we cross-correlated 
 small regions. Since a sharp-edged sample region is very unsuitable
 for this operation, we employed instead a local Gaussian mask with
 an $e^{-1}$ radius of 0.2$\arcsec$.   We measured motions {\it relative
 to the co-moving grid that expands with the lobes,\/}  i.e., a value of
 zero corresponds to ejection date 1850.  The cross-correlation was done
 in both spatial directions $(x, y)$ with shifts ranging from
 $-0.5{\arcsec}$ to $+0.5{\arcsec}$.  The $x, y$ width of each computed
 cell in the correlation datacube was $0.107{\arcsec}$.  We use  
 ``cell'' rather than ``pixel''  to avoid confusion with the
 original image pixels which were much smaller, see \S2.  Because of the
 Gaussian mask mentioned above, the cells are fuzzy and adjoining
 values are not mutually independent.

Figure 6 shows the resulting {\it relative co-moving\/} proper motion\footnote{Hereafter, we 
use motion or  proper motion with the understanding that these are relative motions.} vector field.
At each location the correlation amplitude, $-1$ to $+1$,
indicates the validity of the result;  a motion is considered valid
if the maximum in correlation is greater then 0.4.  As expected, the highest
correlations coincide with strong spatial fine structures, and the
periphery of the butterfly has many correlations above 0.8.
 We experimented with different pairings
 with the  NaCO Ks image and obtained similar results for the high significance correlations. 
The $\sim$ 6 yr interval between the two images and the
$\pm 0.5{\arcsec}$ explored by the correlations pose an upper limit of
$\sim \, 0.01{\arcsec}$/yr per axis on the proper motions.
All of the high-correlation motions are safely below this limit;
e.g., Figure 7 shows one elongated structure with a significant
relative motion of $\sim \, 0.008{\arcsec}$/yr.  The lower limit on
what is measureable depends on how much structure there is in a
given region;  for many regions the 1$\sigma$ limit is 0.1 mas/yr.
We used Monte Carlo simulations to estimate rms uncertainties, assuming
that the error budget is dominated by statistical noise in the
shallower NaCO images, estimated from the pixel-to-pixel 
standard deviation using a robust sigma measurement. For every correlation cell we adopt the
rms error in 100 random trials as the local uncertainty in 
each component of the $x, y$ motion\footnote{Caveat:\/ Obviously
this does not include systematic errors, which 
cannot be quantified with the available information.  Recall,
also, that adjoining cells are not independent.}.  These 
statistical uncertainties are typically 0.1 to 0.2 mas/yr.

There are numerous  knots, blobs and filamentary features in the vector 
map. Not all of them define the bright rims of the butterfly.  For many of
 these we derived specific motions referring to the physical structures
 rather than cells in the $x, y$ grid.  To determine the motions of the more 
 recognizable structures we used a circular virtual aperture with parabolic weight function
 $1 - (r/R)^2$. (Again we must avoid sharp edges.)  For each clear
 physical feature in the map, we first used the virtual aperture to
 measure a local $x, y$ centroid based on brightness.  Then, centering
 the virtual aperture at that position, we refer to this weighted local
 set of cells as an aperture sample or just ``aperture.''
 For each aperture sample we then determined the weighted mean of the motions
 for the included cells.  Experimenting with aperture sizes,
 we found that the results changed little for
 $R \sim 0.16{\arcsec}$;  therefore we adopted a diameter
 $2R = 0.321{\arcsec} = 3$ cell-widths.  
 The dispersions in the mean motions for the cells in the apertures are larger than the 
 weighted quadratic mean of the cells' individual error estimates.
 This may be due to systematic effects. Consequently, we adopted 
  the standard  error of the mean  motions for  the error in each aperture.  
  We retained only aperture samples whose 
 weighted-average correlation amplitudes were greater than 0.8.

The results for the 67 high significance  apertures including the angular distance of each aperture from the star, position angle, and weighted mean relative  motion in R.A. and Dec in the co-moving reference frame with their
 standard errors are in the Appendix.  Since the motions have been measured in the rest-frame 
 of the expanding Homunculus, we have also calculated the intrinsic or true  motions in the fixed
 $x, y$ grid, listed in the last two columns of the Appendix. 
 The distribution of these  apertures on an image of the butterfly is shown
 in Figure 8 with the vectors for the true or intrinsic projected motions. 
The corresponding velocities in the plane of the sky, at a distance of 2.3 kpc for 
$\eta$ Car (Davidson et al 2001, Smith  2002), are given in Table 2 with the epochs 
when the sub-structures, or knots or clumps,  were ejected assuming uniform motion radially 
outward from the star. 
A histogram of the ejection epochs from the projected  motions is shown in Figure 9. 
The names or designations of the larger regions used in Table 2 and shown in Figure 10 
are adopted from Chesneau et al.(2005) and Smith et al. (2003).

Smith (2006)  obtained long-slit spectroscopy of the H$_{2}$ emission line at
2.1$\mu$m across the Homunculus nebula, and three of the slits, SW1, NE1, and ``star'', overlap the
bright rims of the butterfly at five different positions shown in Figure 10.
To determine the total space motion of the knots or sub-structures in these 
apertures, their spatial orientation,  and therefore that of the bright
rims of the butterfly, we used the corresponding radial (Doppler) velocities,  
 measured relative to the Doppler velocity of $\eta$ Car, combined with the projected velocity from the proper motions at these five positions.
The radial velocities for the apertures within the slits, their resulting total velocity relative to the star, 
the orientation angle relative to the plane of the sky, $\theta$, linear distance from the star, and the corresponding ejection 
epoch are all included in  Table 2 for the relevant apertures.  

    Throughout this paper we assume that the ejecta move radially  
        outward from the central star, and that any apparent transverse 
	    motions are measurement errors.  Angular momentum considerations  
	        make this a safe assumption;  given the observed total speeds, 
		    off-center ejection in $\eta$ Car's binary orbit would lead to 
		        transverse motions less than 0.1 mas yr$^{-1}$ at 
			    $r \sim 3000$ AU.  

\section{Discussion -- Expansion, Age,  and Distribution of the Inner Ejecta} 

It is perhaps not surprising that the ejection epochs cluster around two dates, circa 
 $\eta$ Car's great eruption (1838 -- 1858) and near 1900 corresponding to its second eruption 
(1887 -- 1895); see Figure 9. Like many previous authors we find a date {\it after\/} 1895. However, acceleration
at the 10\% level is physically possible and may explain this discrepancy as noted by Davidson et al. (1997) (see also Smith et al. 2004).\footnote{Note also that in principle the emission center of a condensation could migrate 
     outward faster than its center of mass, because the spatial distribution 
          of excitation depends on the local UV flux and gas density. This effect 
	       is small, however, for any brightness peak that represents a real 
	            density maximum.} 

The orientation angle ($\theta$) relative to the plane of the sky has a wide  range from  
positive to negative. Most of
the sub-regions in Table 2, however, are projected away from us at angles from $\approx +20\arcdeg$ to $+60\arcdeg$.
The equatorial plane (see the schematic diagram in Figure 11) is tilted away from us,
 behind the SE lobe, and toward us, projected in front of the NW lobe, at about 41$\arcdeg$.
The clumps of knots moving away from us in the SE lobe thus appear to be distributed  above 
and below the equatorial plane. Either the equatorial debris has a fairly wide opening angle of $\approx$ 
20$\arcdeg$ or the material was ejected at somewhat different latitudes relative to the plane.
Numerous knots or apertures within the larger regions have comparable $\theta$'s 
suggesting that the different projection angles are significant.
There may also be some correlation between the ejection epoch and the orientation angle, although 
our information for $\theta$ is far from complete.  For example, all of the 
knots with $\theta$ near 20$\arcdeg$ (S. Clump and S. Arc) and those with $\theta$ 
near  50$\arcdeg$ (SE Clump) have ejection dates corresponding
to the second eruption, while those consistent with the great eruption have orientation 
angles closer to  40$\arcdeg$.  Four of the sub-regions,  apertures 3, 65, 66,
and 67 in the ``SW Region'' and the ``NE region'',  are oriented towards us and are very likely
 due to material projected in front of the NW lobe  ejected at the time of the 
 great eruption.

The total space motions or velocities measured for several of the bright clumps or 
knots range from about 60 to 100 km s$^{-1}$, much slower that the polar expansion of
either the Homunculus at $\approx$ 600  km s$^{-1}$ (see references in the Introduction)\footnote{There have been numerous measurements of the expansion of the Homunculus. See review by 
Davidson and Humphreys (1997) and references therein for the earlier work.} or the little 
Homunculus at 300  km s$^{-1}$ (Ishibashi et al. 2003, see also Smith 2005).  
However, material associated with the  equatorial plane 
has been found to be expanding 
at a range of  velocities \citep{KD01} including relatively slow moving ejecta near the 
star. Davidson et al. (2001) also concluded that that the equatorial debris consisted of 
material from the two eruption events. Zethson et al. (1999) reported slow moving gas 
projected in front of the NW lobe with radial velocities relative
to the star of $-40$ to $-140$ km s$^{-1}$, and presumably in the 
equatorial plane. The Weigelt knots, 
mentioned earlier, near the equatorial plane,  are also slow moving, with radial 
velocities of $\approx$ $-50$ km s$^{-1}$ and an estimated ejection epoch of 1900.

 Our results for the bright clumps and knots that define the apparent rim of 
the butterfly are thus consistent with results from previous work. They are  slow moving, from
two different eruptions, and  near the equatorial plane, but projected away from the plane of the sky, $\approx$  2000 to 4000 AU from the star.   In
Figure 12 we show the physical location of the separate clumps and filamentary
features relative to the equatorial plane on a schematic diagram of the inner region of the Homunculus. 
The latitude or angle relative to the equatorial plane, $\zeta$, is also included in Table 2. 
Given the orientation and distances  of the sub-regions from the star, most of them are
on the far side of the Little Homunculus and are also very likely  on the far side of the SE lobe.  Figure 8 in Ishibashi et al. (2003) llustrates the complexity  
of this region,  and  also shows a cloud of debris within 1$\arcsec$ of the star beyond the 
apparent boundaries of the Little Homunculus and  bracketing the equatorial plane, similar 
to the spatial distribution of the sub-regions or clumps in the butterfly.  

Smith (2006) suggested that the butterfly is part of 
a ``disrupted torus,'' related to a circular region where the two 
Homunculus lobes meet near the configuration's midplane.  This idea 
can probably  be modified to be fairly consistent with the ejecta directions 
and two ejection epochs described above. 
On the other hand, our results are at least equally consistent 
with other possibilities.   Modern investigations (including the work 
here) reveal complex asymmetry, not circular structure, in and around 
the equatorial plane where the lobes would meet in a simple picture.  
This is not very surprising, since there is no strong theoretical 
reason to expect either circular or latitudinal symmetry there. 
The central binary system is generally acknowledged to 
have a very eccentric orbit which obviously breaks the  
circular symmetry for outflows near periastron;  
in principle the star's equatorial region is vulnerable to 
instabilities that may cause local and chaotic low-speed ejections   
     (e.g., see remarks in \S 5 of \citep{Zeth99}).  
Furthermore, details of the large-scale Homunculus lobe shapes and orientations 
     show that 
     axial symmetry was only an approximation for the great eruption  
              \citep{Morse1998,KD01}, and   
    large-scale radial ``streamers'' or ``jets'' well beyond the lobes have 
    long been known in the images  
 for some time  (e.g., (Duschl et al. 1995, Meaburn et al. 1987, 1993, 1996).  
 Given these diverse clues, we should not expect to represent the past or 
 present equatorial region of the Homunculus in terms of two well-defined, 
  intersecting, axially symmetric geometrical lobe-figures. Instead, 
 the lobes appear to become ill-defined near the midplane.  A large 
  circular configuration  around $\eta$ Car is worth considering, 
  but there is no strong observational or theoretical 
 evidence that such a structure exists or existed. Note that the famous ``equatorial 
 skirt'' -- located outside the inner region discussed here -- consists of radial structures with no discernible circular symmetry. 

In the absence of quantitative
dynamical models, perhaps the {\it simplest} hypothesis is that either the unstable primary star
 can can eject matter asymmetrically, and/or  
  ejections can occur near periastron when the companion star's 
 gravity and wind perturb the outflows asymmetrically.  These 
 phenomena would explain our observations (qualitatively, at least) 
 wihout any concentration of pre-ejected material near the 
  midplane. Moreover, in either case, local flows can be erratic and 
  asymmetric relative to the midplane because they are very 
 sensitive to unstable local conditions.  These possibilities  
  were implicit in some of the references cited above, but they 
 are extremely difficult to model physically.

In summary, the appearance of the ``butterfly nebula'' as a single, coherent structure with multiple lobes
or wings or as an  equatorial torus, was thus due to  projection effects and to low resolution in the earlier studies. Instead the knots or
clumps of knots are separate, ejected primarily at two different epochs,  
and are apparently oriented within $\pm 10\arcdeg$ to $\pm 25\arcdeg$ of the 
equatorial plane, $\sim$  2000 to 4000 AU from the star. They may be filamentary streamers or jet-like features associated 
with the equatorial debris on the far side of the SE lobe.

\acknowledgments
We gratefully acknowledge the support of the Gemini staff and the Director's
office for granting Direcor's Discretionary time for these observations. 
Chesneau thanks Ph.~Bendjoya, G.~Guerri and A. Collioud for fruitful
discussions and testing of analysis methods of the NaCO data.
When this work was begun,  Artigau was supported by the Gemini Observatory, 
which is operated by the Association of Universities
for Research in Astronomy, Inc., on behalf of the international Gemini partnership of Argentina,
Australia, Brazil, Canada, Chile, the United Kingdom, and the United States of America.

\appendix 
\section{Positions and Projected  Motions for the Apertures}


\begin{deluxetable}{lccc}
\tablewidth{0pt}
\tabletypesize{\scriptsize}
\tablecaption{NICI Filter Selection}
\tablecolumns{4}
\tablehead{ \colhead{Filter}& \colhead{Channel}&
  \colhead{Central $\lambda$ ($\mu$m)}&\colhead{Width ($\mu$m)}
}
\startdata
[Fe II]&Blue&1.644&0.0247\\
CH4 H 1\% L&Red&1.628&0.0174\\
H2 1-0 S(1)&Blue&2.124&0.0261\\
Br-gamma&Blue&2.169&0.0295\\
Kcont&Red&2.272&0.0352\\
\enddata
\end{deluxetable}

\begin{deluxetable}{ccccccccc}   
  \tablewidth{0pt}
  \tabletypesize{\scriptsize}
  \tablecaption{Velocities and Ejection Epochs \label{datatab}}
  \tablecolumns{9}
  \tablehead{
    \colhead{Region}&
    \colhead{Projected Vel}&
    \colhead{Origin}&
    \colhead{Doppler Vel}&
    \colhead{Total Vel}&
    \colhead{$\theta$\tablenotemark{b}}&
    \colhead{Distance\tablenotemark{d}}&
    \colhead{Origin}&  
    \colhead{$\zeta$\tablenotemark{e}}\\
    \colhead{Ap. Number}&
    \colhead{(km s$^{-1}$)}&
    \colhead{Epoch\tablenotemark{a}}&
    \colhead{(km s$^{-1}$)}&
    \colhead{(km s$^{-1}$)}&
    \colhead{(deg)}&
    \colhead{(AU)}&
    \colhead{Epoch\tablenotemark{c}}&  
    \colhead{(deg)}
  }
  \startdata
\multicolumn{9}{l}{A:  Western Arc}\\
1&87.1$\pm$3.4&1902$\pm$8\\
2&57.2$\pm$5.8&1830$\pm$36\\
\multicolumn{9}{l}{B:  SW Region}\\
3&81.2$\pm$5.2&1848$\pm$20&-52$\pm$5&96.7$\pm$6.0&$-$32.9$\pm$3.4&3286&1845$\pm$17&$-$2\\
4&145.7$\pm$3.7&1906$\pm$5\\
5&145.4$\pm$2.2&1911$\pm$3\\
6&159.8$\pm$2.6&1917$\pm$3\\
7&164.2$\pm$2.5&1903$\pm$3\\
8&144.2$\pm$2.0&1877$\pm$4\\
9&138.2$\pm$1.9&1865$\pm$4\\
10&122.7$\pm$1.1&1852$\pm$3\\
11&155.4$\pm$1.3&1871$\pm$2\\
12&184.6$\pm$2.2&1883$\pm$3\\
\multicolumn{9}{l}{C:  S Clump}\\
13&101.1$\pm$2.0&1854$\pm$6\\
14&100.2$\pm$1.8&1885$\pm$4&38$\pm$5&107.3$\pm$5.3&21.0$\pm$3.7&2807&1883$\pm$10&$-$4\\
15&91.0$\pm$3.5&1902$\pm$8\\
16&68.9$\pm$2.6&1901$\pm$8\\
17&96.5$\pm$1.4&1898$\pm$3\\
18&119.4$\pm$2.0&1897$\pm$4&38$\pm$5&125.3$\pm$5.4&17.7$\pm$3.5&2945&1896$\pm$8&$-$8\\
\multicolumn{9}{l}{D:  Southern Arc}\\
19&116.7$\pm$2.0&1898$\pm$4&38$\pm$5&122.7$\pm$5.4&18.0$\pm$3.5&2877&1896$\pm$8&$-$13\\
20&135.7$\pm$1.7&1907$\pm$3&38$\pm$5&140.8$\pm$5.3&15.6$\pm$3.2&3032&1905$\pm$7&$-$13\\
21&150.3$\pm$1.2&1910$\pm$2&38$\pm$5&155.0$\pm$5.1&14.9$\pm$3.0&3236&1908$\pm$6&$-$17\\
22&103.2$\pm$0.7&1887$\pm$2\\
23&98.2$\pm$0.8&1895$\pm$2\\
24&86.4$\pm$1.6&1887$\pm$4\\
25&77.5$\pm$1.9&1900$\pm$5\\
\multicolumn{9}{l}{E:  SE Clump and Filament}\\
26&61.1$\pm$1.1&1901$\pm$4\\
27&44.3$\pm$1.7&1892$\pm$9\\
28&29.2$\pm$1.6&1878$\pm$14&48$\pm$5&56.2$\pm$5.3&58.7$\pm$2.5&1549&1876$\pm$14&$+$18\\
29&33.6$\pm$2.2&1881$\pm$16&48$\pm$5&58.6$\pm$5.4&55.1$\pm$3.0&1608&1877$\pm$15&$+$17\\
30&39.9$\pm$1.9&1874$\pm$12&48$\pm$5&62.4$\pm$5.3&50.0$\pm$2.9&1796&1870$\pm$14&$+$8\\
31&43.8$\pm$2.2&1879$\pm$12&48$\pm$5&65.0$\pm$5.4&47.6$\pm$3.1&1773&1878$\pm$13&$+$3\\
32&52.6$\pm$1.4&1886$\pm$7&48$\pm$5&71.2$\pm$5.2&42.4$\pm$2.9&1837&1885$\pm$11&$-$1\\
33&52.3$\pm$1.9&1862$\pm$11&48$\pm$5&71.0$\pm$5.4&42.5$\pm$3.1&2214&1859$\pm$14&$-$1\\
\multicolumn{9}{l}{F: SE Arc}\\
34&57.5$\pm$1.4&1869$\pm$7&48$\pm$5&74.9$\pm$5.2&39.9$\pm$2.9&2218&1866$\pm$13&$-$12\\
35&63.3$\pm$1.3&1848$\pm$7&48$\pm$5&79.4$\pm$5.2&37.2$\pm$3.0&2685&1846$\pm$14&$-$1\\
36&70.4$\pm$1.7&1852$\pm$7&48$\pm$5&85.2$\pm$5.3&34.3$\pm$3.2&2811&1850$\pm$13&$-$7\\
37&80.7$\pm$1.5&1857$\pm$6&48$\pm$5&93.9$\pm$5.2&30.7$\pm$3.1&3022&1854$\pm$12&$-$13\\
38&80.2$\pm$2.9&1825$\pm$13&48$\pm$5&93.5$\pm$5.8&30.9$\pm$3.6&3618&1822$\pm$16&$-$-11\\
39&82.9$\pm$3.0&1817$\pm$14&48$\pm$5&95.8$\pm$5.8&30.1$\pm$3.6&3880&1814$\pm$17&$-$5\\
\multicolumn{9}{l}{G: East Edge Connecting SE Arc and NE Region}\\
40&95.0$\pm$1.1&1838$\pm$4\\
\multicolumn{9}{l}{H:  NE Clump}\\
41&116.1$\pm$1.1&1863$\pm$3\\
42&78.8$\pm$1.4&1830$\pm$6&58$\pm$5&97.8$\pm$5.2&36.4$\pm$2.5&3685&1827$\pm$13&$+$10\\
43&85.3$\pm$2.5&1862$\pm$8&58$\pm$5&103.2$\pm$5.6&34.2$\pm$2.9&3199&1860$\pm$11&$+$10\\
44&60.5$\pm$2.6&1817$\pm$16&58$\pm$5&83.8$\pm$5.6&43.8$\pm$2.9&3377&1815$\pm$16&$+$14\\
45&58.5$\pm$1.1&1839$\pm$6&58$\pm$5&82.4$\pm$5.1&44.8$\pm$2.2&2949&1837$\pm$13&$+$19\\
46&58.4$\pm$0.9&1843$\pm$5&58$\pm$5&82.3$\pm$5.1&44.8$\pm$2.1&2884&1840$\pm$13&$+$22\\
47&61.9$\pm$4.3&1867$\pm$19\\
48&42.1$\pm$6.0&1827$\pm$51\\
\multicolumn{9}{l}{I:  South side of NE Region}\\
49&86.2$\pm$1.7&1848$\pm$6\\
50&74.7$\pm$1.1&1832$\pm$5&58$\pm$5&94.5$\pm$5.1&37.8$\pm$2.4&3521&1829$\pm$13&$+$23\\
51&74.5$\pm$1.3&1833$\pm$6&58$\pm$5&94.4$\pm$5.2&37.9$\pm$2.4&3497&1830$\pm$13&$+$19\\
52&82.9$\pm$1.7&1829$\pm$7\\
53&77.5$\pm$1.2&1802$\pm$6\\
54&120.9$\pm$1.8&1860$\pm$4\\
55&125.8$\pm$2.0&1845$\pm$5\\
56&128.8$\pm$1.3&1840$\pm$3\\
57&153.2$\pm$3.0&1877$\pm$5\\
\multicolumn{9}{l}{J:  North side of NE Region}\\
58&110.3$\pm$2.9&1844$\pm$8\\
59&128.3$\pm$2.3&1852$\pm$6\\
60&107.3$\pm$2.5&1843$\pm$7\\
61&94.4$\pm$2.2&1835$\pm$8\\
62&83.3$\pm$2.9&1831$\pm$12\\
63&111.7$\pm$1.2&1861$\pm$3\\
64&91.3$\pm$0.7&1843$\pm$2\\
65&85.3$\pm$1.6&1850$\pm$6&-62$\pm$5&105.5$\pm$5.3&$-$36.0$\pm$2.4&3524&1848$\pm$14&$-$14\\
66&87.9$\pm$2.0&1862$\pm$6&-62$\pm$5&107.6$\pm$5.4&$-$35.1$\pm$2.5&3316&1860$\pm$13&$-$12\\
67&96.7$\pm$5.0&1869$\pm$14&-62$\pm$5&114.9$\pm$7.1&$-$32.7$\pm$3.6&3388&1867$\pm$16&$-$8\\
\enddata
\tablenotetext{a}{From the total projected motion in the plane of the sky.}
\tablenotetext{b}{Orientation angle with respect to the plane of the sky.}
\tablenotetext{c}{From the total space velocity.}
\tablenotetext{d}{Distance from the star in AU at the inclination angle.}
\tablenotetext{e}{The angle or latitude relative to the equatorial plane.}
\end{deluxetable}

\begin{deluxetable}{ccrrrrrrrr}   
\tablewidth{0pt}
\tabletypesize{\scriptsize}
\tablenum{A}
\tablecaption{Positions and  Proper Motions \label{appendixtab}}
\tablecolumns{10}
\tablehead{
\colhead{Aperture}&
\colhead{Dist. from}&
\colhead{PA\tablenotemark{a}}&
 \multicolumn{2}{c}{Relative Proper Motion}&
  &
\colhead{Avg}&
\colhead{Homunculus\tablenotemark{d}}&
\multicolumn{2}{c}{True Motion\tablenotemark{e}}
    \\
\colhead{Number}&
\colhead{Star (arcsec)}&
\colhead{(deg)}&
\colhead{RA (mas/yr)}&
\colhead{Dec (mas/yr)}&
 \colhead{N\tablenotemark{b}}&
  \colhead{Sig\tablenotemark{c}}&
 \colhead{Exp. (mas/yr)}&
 \colhead{RA (mas/yr)}&
 \colhead{Dec (mas/yr)}
 }
\startdata
1&0.86&-77.29&-2.56$\pm$0.58&0.46$\pm$0.17&8&0.93&5.39&-7.82$\pm$0.58&1.64$\pm$0.17\\
2&0.94&-80.21&0.60$\pm$0.78&-0.63$\pm$0.69&6&0.93&5.92&-5.23$\pm$0.78&0.38$\pm$0.69\\
3&1.20&-99.73&0.77$\pm$0.12&-2.09$\pm$0.92&8&0.84&7.53&-6.65$\pm$0.12&-3.36$\pm$0.92\\
4&1.38&-124.15&-2.78$\pm$0.35&-4.01$\pm$0.56&7&0.90&8.70&-9.98$\pm$0.35&-8.89$\pm$0.56\\
5&1.30&-127.36&-2.45$\pm$0.13&-4.90$\pm$0.38&7&0.92&8.19&-8.96$\pm$0.13&-9.88$\pm$0.38\\
6&1.35&-130.02&-2.98$\pm$0.18&-5.71$\pm$0.43&7&0.91&8.50&-9.49$\pm$0.18&-11.17$\pm$0.43\\
7&1.59&-142.85&-3.43$\pm$0.37&-3.71$\pm$0.25&7&0.89&10.03&-9.48$\pm$0.37&-11.71$\pm$0.25\\
8&1.75&-147.30&0.05$\pm$0.15&-2.58$\pm$0.33&8&0.92&11.00&-5.90$\pm$0.15&-11.84$\pm$0.33\\
9&1.83&-149.59&0.40$\pm$0.20&-1.53$\pm$0.27&7&0.91&11.52&-5.43$\pm$0.20&-11.46$\pm$0.27\\
10&1.77&-153.81&0.15$\pm$0.12&-0.20$\pm$0.16&7&0.89&11.15&-4.77$\pm$0.12&-10.20$\pm$0.16\\
11&1.97&-157.71&-0.34$\pm$0.10&-1.90$\pm$0.20&9&0.91&12.37&-5.03$\pm$0.10&-13.34$\pm$0.20\\
12&2.14&-161.66&-3.36$\pm$0.38&-2.36$\pm$0.07&7&0.85&13.46&-7.59$\pm$0.38&-15.14$\pm$0.07\\
13&1.44&-165.25&0.42$\pm$0.17&-0.34$\pm$0.33&7&0.90&9.04&-1.88$\pm$0.17&-9.08$\pm$0.33\\
14&1.14&-166.95&0.13$\pm$0.05&-2.09$\pm$0.31&7&0.93&7.16&-1.49$\pm$0.05&-9.07$\pm$0.31\\
15&0.89&-172.76&1.21$\pm$0.10&-2.77$\pm$0.62&7&0.94&5.60&0.51$\pm$0.10&-8.33$\pm$0.62\\
16&0.68&-176.65&-0.53$\pm$0.41&-1.98$\pm$0.23&6&0.61&4.30&-0.78$\pm$0.41&-6.27$\pm$0.23\\
17&0.99&-172.40&0.78$\pm$0.16&-2.71$\pm$0.19&8&0.93&6.20&-0.04$\pm$0.16&-8.85$\pm$0.19\\
18&1.22&-169.59&-0.36$\pm$0.14&-3.24$\pm$0.33&7&0.94&7.69&-1.75$\pm$0.14&-10.81$\pm$0.33\\
19&1.19&-178.95&-0.66$\pm$0.17&-3.20$\pm$0.33&8&0.92&7.48&-0.79$\pm$0.17&-10.68$\pm$0.33\\
20&1.27&-175.76&-1.16$\pm$0.10&-4.33$\pm$0.29&8&0.93&8.02&-1.75$\pm$0.10&-12.33$\pm$0.29\\
21&1.36&178.64&-1.24$\pm$0.08&-5.17$\pm$0.20&7&0.92&8.58&-1.03$\pm$0.08&-13.75$\pm$0.20\\
22&1.16&172.11&0.24$\pm$0.07&-2.17$\pm$0.09&6&0.93&7.30&1.24$\pm$0.07&-9.39$\pm$0.09\\
23&1.03&172.10&0.56$\pm$0.04&-2.50$\pm$0.14&6&0.92&6.45&1.44$\pm$0.04&-8.89$\pm$0.14\\
24&0.97&165.09&0.58$\pm$0.13&-1.73$\pm$0.26&7&0.89&6.10&2.15$\pm$0.13&-7.63$\pm$0.26\\
25&0.78&164.06&0.34$\pm$0.24&-2.20$\pm$0.23&9&0.88&4.89&1.68$\pm$0.24&-6.91$\pm$0.23\\
26&0.61&161.24&-0.83$\pm$0.12&-1.97$\pm$0.15&7&0.92&3.82&0.40$\pm$0.12&-5.59$\pm$0.15\\
27&0.48&158.17&-0.59$\pm$0.08&-1.25$\pm$0.29&7&0.96&3.00&0.53$\pm$0.08&-4.03$\pm$0.29\\
28&0.35&156.25&-0.97$\pm$0.24&-0.67$\pm$0.16&7&0.91&2.20&-0.08$\pm$0.24&-2.68$\pm$0.16\\
29&0.40&140.81&-1.64$\pm$0.37&-1.15$\pm$0.11&7&0.87&2.49&-0.07$\pm$0.37&-3.08$\pm$0.11\\
30&0.50&136.63&-0.95$\pm$0.26&-1.19$\pm$0.22&7&0.91&3.12&1.19$\pm$0.26&-3.46$\pm$0.22\\
31&0.52&120.74&0.80$\pm$0.34&-0.06$\pm$0.18&6&0.87&3.28&3.62$\pm$0.34&-1.74$\pm$0.18\\
32&0.59&118.35&1.28$\pm$0.19&0.20$\pm$0.17&7&0.90&3.73&4.56$\pm$0.19&-1.57$\pm$0.17\\
33&0.71&113.46&0.38$\pm$0.31&-0.01$\pm$0.15&7&0.86&4.44&4.46$\pm$0.31&-1.78$\pm$0.15\\
34&0.74&135.01&1.30$\pm$0.23&-1.61$\pm$0.10&8&0.95&4.64&1.98$\pm$0.23&-4.89$\pm$0.10\\
35&0.93&131.20&-3.24$\pm$0.22&-1.82$\pm$0.10&7&0.96&5.87&1.18$\pm$0.22&-5.69$\pm$0.10\\
36&1.01&130.18&-2.65$\pm$0.24&-1.96$\pm$0.19&7&0.94&6.38&2.22$\pm$0.24&-6.07$\pm$0.19\\
37&1.13&129.86&-1.59$\pm$0.24&-1.77$\pm$0.15&7&0.90&7.09&3.85$\pm$0.24&-6.32$\pm$0.15\\
38&1.35&129.26&-3.49$\pm$0.51&-1.30$\pm$0.10&7&0.89&8.50&3.09$\pm$0.51&-6.68$\pm$0.10\\
39&1.46&126.55&-6.47$\pm$0.51&-2.08$\pm$0.16&8&0.86&9.18&0.91$\pm$0.51&-7.55$\pm$0.16\\
40&1.49&100.34&-0.51$\pm$0.08&1.01$\pm$0.17&6&0.84&9.35&8.69$\pm$0.08&-0.67$\pm$0.17\\
41&1.55&70.35&0.70$\pm$0.20&0.66$\pm$0.06&7&0.87&9.77&9.90$\pm$0.20&3.94$\pm$0.06\\
42&1.29&66.91&-0.72$\pm$0.23&-0.66$\pm$0.09&7&0.92&8.14&6.77$\pm$0.23&2.53$\pm$0.09\\
43&1.15&65.23&0.92$\pm$0.42&-0.79$\pm$0.16&6&0.93&7.25&7.50$\pm$0.42&2.24$\pm$0.16\\
44&1.06&71.62&-0.82$\pm$0.46&-1.58$\pm$0.09&7&0.80&6.69&5.53$\pm$0.46&0.53$\pm$0.09\\
45&0.91&62.92&0.07$\pm$0.14&-1.17$\pm$0.13&7&0.96&5.73&5.17$\pm$0.14&1.44$\pm$0.13\\
46&0.89&56.25&0.19$\pm$0.14&-0.80$\pm$0.06&7&0.95&5.59&4.84$\pm$0.14&2.30$\pm$0.06\\
47&0.81&60.32&-0.21$\pm$0.10&1.29$\pm$0.76&7&0.83&5.08&4.21$\pm$0.10&3.81$\pm$0.76\\
48&0.70&45.56&0.17$\pm$0.41&-1.14$\pm$1.00&7&0.96&4.43&3.33$\pm$0.41&1.96$\pm$1.00\\
49&1.27&42.79&1.71$\pm$0.19&-2.49$\pm$0.24&7&0.86&8.01&7.15$\pm$0.19&3.39$\pm$0.24\\
50&1.21&46.76&0.38$\pm$0.19&-1.82$\pm$0.08&8&0.86&7.63&5.94$\pm$0.19&3.41$\pm$0.08\\
51&1.20&53.96&-0.02$\pm$0.22&-1.36$\pm$0.07&7&0.87&7.58&6.10$\pm$0.22&3.10$\pm$0.07\\
52&1.37&57.44&-0.70$\pm$0.20&-0.81$\pm$0.24&8&0.94&8.62&6.57$\pm$0.20&3.83$\pm$0.24\\
53&1.48&53.85&-2.23$\pm$0.18&-0.69$\pm$0.12&8&0.89&9.28&5.26$\pm$0.18&4.79$\pm$0.12\\
54&1.66&52.73&-0.21$\pm$0.23&1.29$\pm$0.23&7&0.83&10.41&8.08$\pm$0.23&7.60$\pm$0.23\\
55&1.89&59.35&-0.74$\pm$0.28&0.51$\pm$0.21&7&0.83&11.88&9.49$\pm$0.28&6.57$\pm$0.21\\
56&2.00&60.32&-1.02$\pm$0.19&0.25$\pm$0.15&8&0.85&12.55&9.89$\pm$0.19&6.47$\pm$0.15\\
57&1.86&51.14&2.70$\pm$0.48&0.33$\pm$0.21&6&0.85&11.67&11.79$\pm$0.48&7.66$\pm$0.22\\
58&1.67&38.10&0.47$\pm$0.16&-0.88$\pm$0.49&7&0.85&10.48&6.94$\pm$0.16&7.37$\pm$0.49\\
59&1.85&37.22&0.89$\pm$0.30&-0.56$\pm$0.29&7&0.84&11.63&7.93$\pm$0.30&8.70$\pm$0.29\\
60&1.65&33.17&-0.68$\pm$0.24&-0.12$\pm$0.37&7&0.88&10.36&4.99$\pm$0.24&8.55$\pm$0.37\\
61&1.51&32.52&-0.94$\pm$0.23&-0.40$\pm$0.33&7&0.92&9.49&4.16$\pm$0.23&7.60$\pm$0.33\\
62&1.36&32.07&-0.33$\pm$0.33&-0.85$\pm$0.42&8&0.93&8.53&4.21$\pm$0.33&6.38$\pm$0.42\\
63&1.52&24.55&-1.40$\pm$0.16&1.24$\pm$0.15&8&0.91&9.54&2.56$\pm$0.16&9.92$\pm$0.15\\
64&1.39&26.87&-1.63$\pm$0.08&0.23$\pm$0.10&7&0.92&8.76&2.33$\pm$0.08&8.05$\pm$0.10\\
65&1.24&24.87&-0.74$\pm$0.27&0.32$\pm$0.13&7&0.94&7.80&2.54$\pm$0.27&7.40$\pm$0.13\\
66&1.18&22.27&0.05$\pm$0.28&0.64$\pm$0.22&8&0.93&7.45&2.88$\pm$0.28&7.53$\pm$0.22\\
67&1.24&18.51&0.18$\pm$0.31&1.05$\pm$0.84&7&0.92&7.82&2.66$\pm$0.31&8.46$\pm$0.84\\
  \enddata
\tablenotetext{a}{Position angle measured from north through east.}
\tablenotetext{b}{Number of cells included in the aperture.}
\tablenotetext{c}{The weighted average of the significance of the cells included in
the aperture.}
\tablenotetext{d}{The radial expansion of the Homunculus at this position from the star.}
\tablenotetext{e}{The relative motion corrected for the expansion of the Homunculus.}
\end{deluxetable}


\begin{figure}
\figurenum{1}
\epsscale{1.0}
\plotone{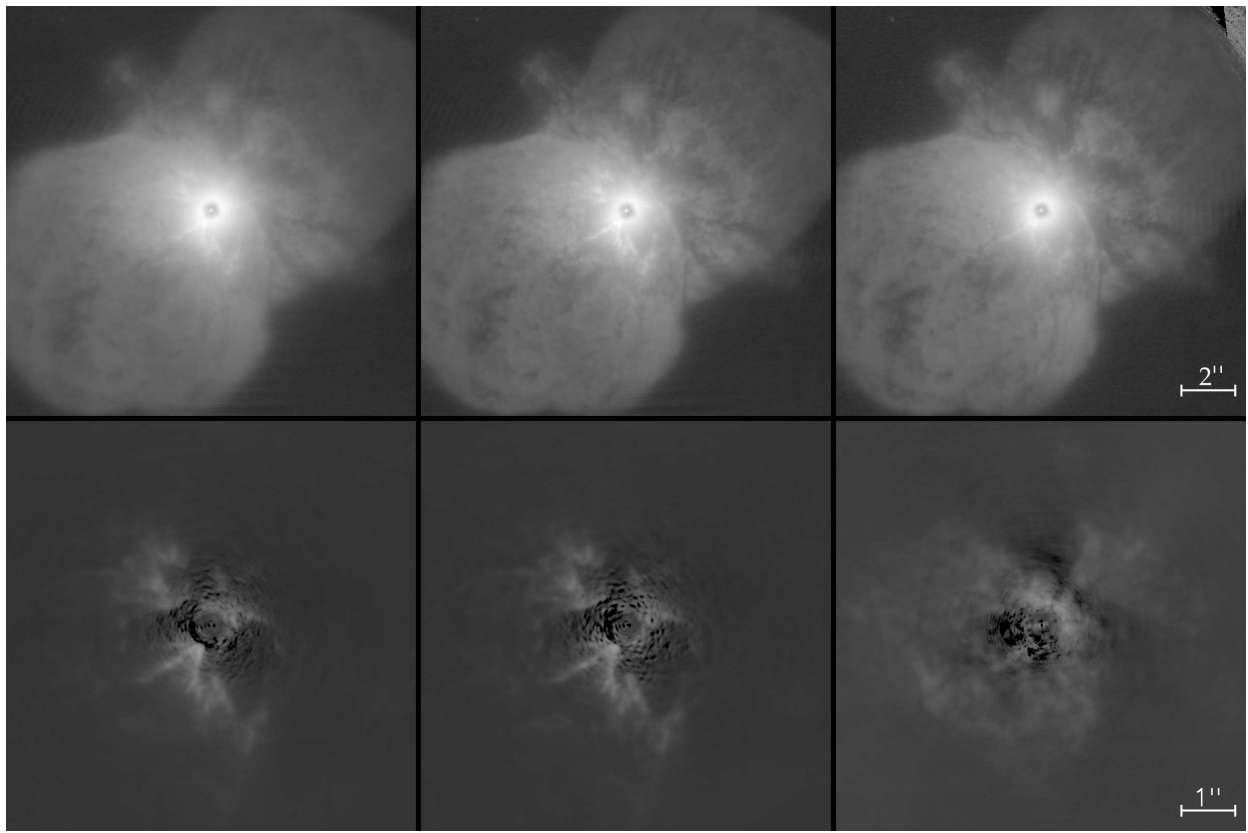} 
\caption{The continuum (top row) and continuum-subtracted (bottom row) H$_{2}$, Br$\gamma$, and [Fe II] images. The grey scale is arbitrary. All of the images are oriented north at the top and east to the left.} 
\end{figure} 

\begin{figure} 
\figurenum{2a} 
\epsscale{1.0} 
\plotone{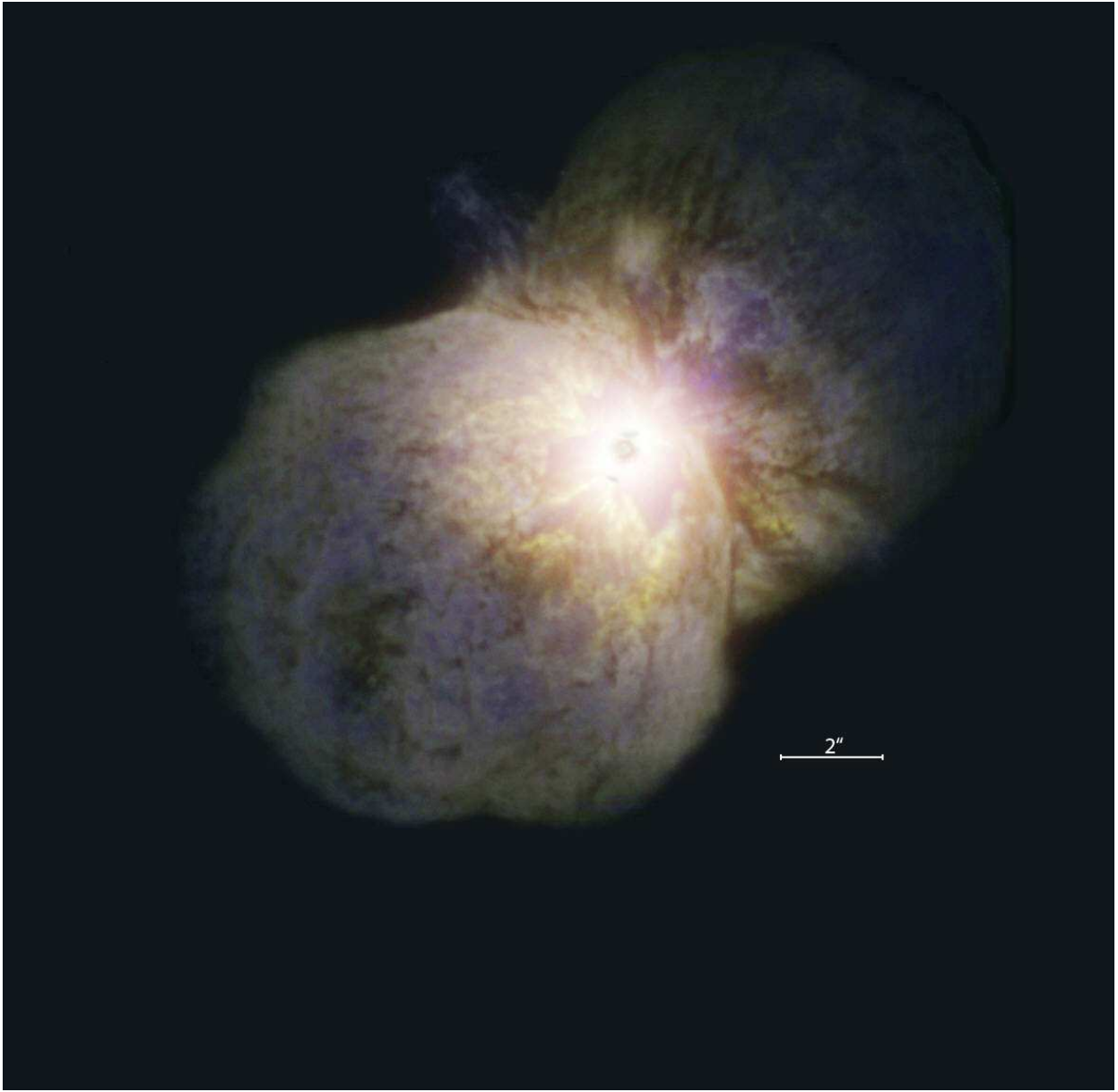} 
\caption{The combined Br$\gamma$, H$_{2}$, [Fe II] image without the continuum subtraction; Br$\gamma$ (red), H$_{2}$ (green),and [Fe II] (blue). North is up and east is left.} 
\end{figure} 

\begin{figure} 
\figurenum{2b} 
\epsscale{0.8} 
\plotone{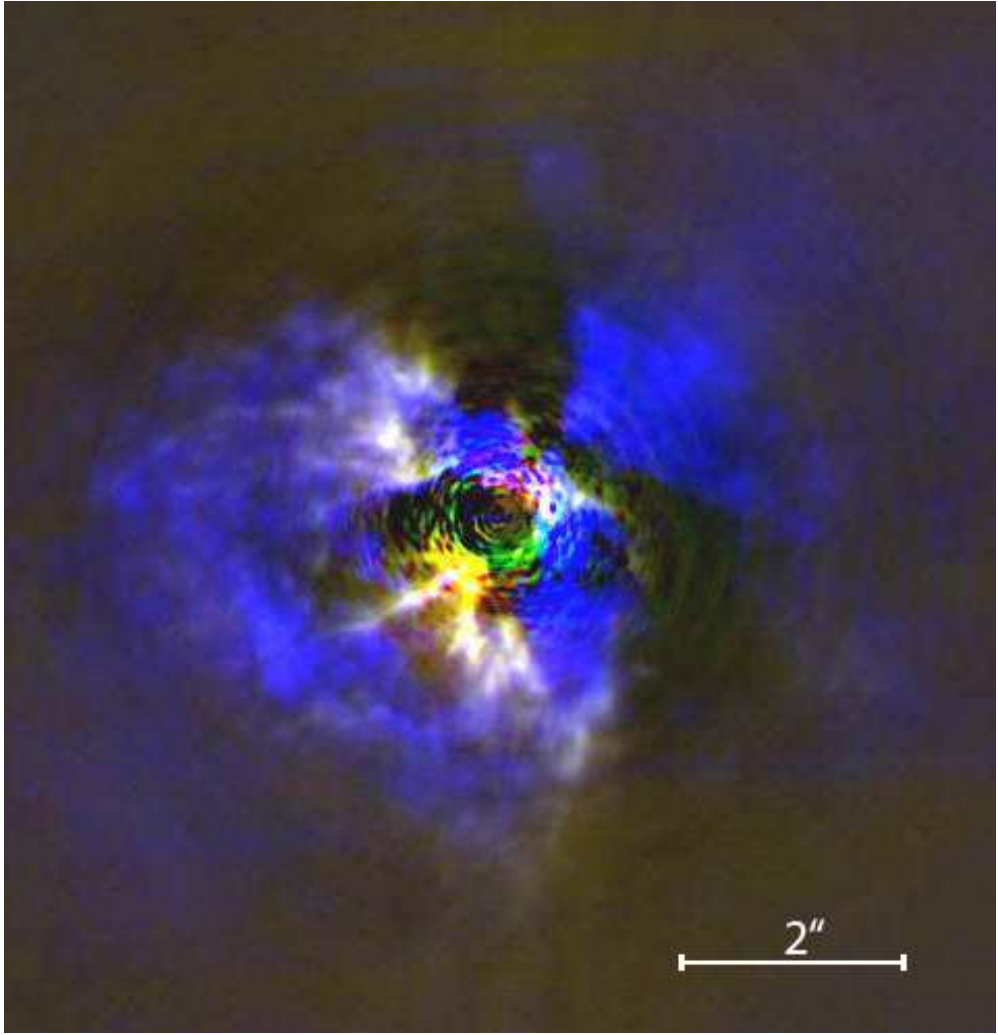} 
\caption{The combined line - continuum image; Br$\gamma$ (red), H$_{2}$ (green),and [Fe II] (blue). This figure shows the inner part of the Homunculus and region of the butterfly. Note the very different distribution of the [Fe II] emission  compared with Br$\gamma$ and H$_{2}$. Br$\gamma$ and H$_{2}$ have nearly identical distributions, therefore their combined image appears white or yellowish.} 
\end{figure} 

\begin{figure} 
\figurenum{3} 
\epsscale{1.0} 
\plotone{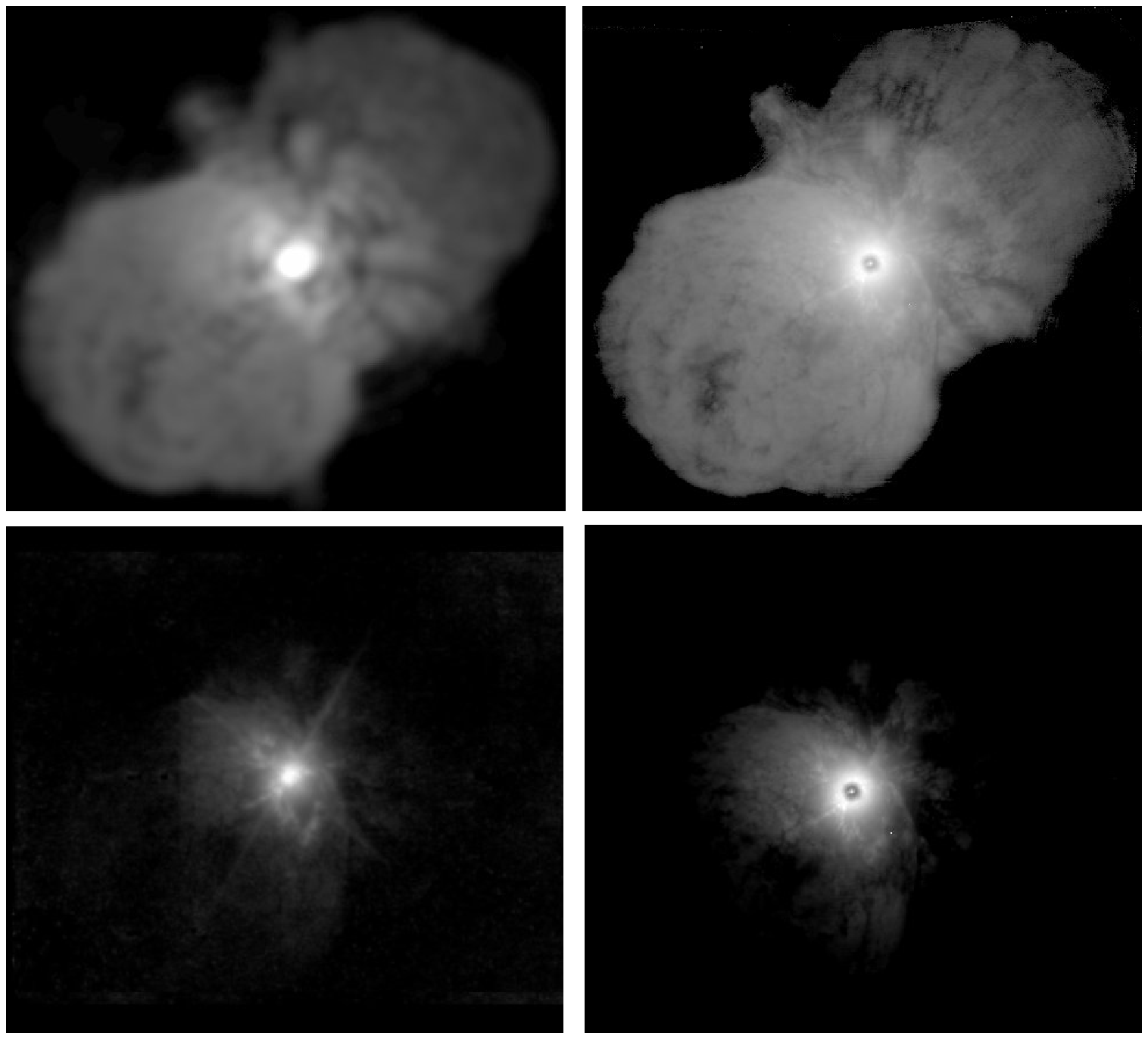} 
\caption{The NICMOS 2.15$\mu$m \citep{Smith00} (top left) compared with the  NICI $K$-band continuum image (top right) and on the bottom row the central region from the 
NaCO broadband K$_{\it s}$ \citep{Ches05} (left) and the NICI $K$-band images (right). 
All four images have been matched to the same spatial scale and orientation, north at the top and east to the left. 
The field of view, 16.5$\arcsec$, is the same.  A different contrast and intensity scaling factor has been used for the upper and lower NICI images}
\end{figure}

\begin{figure}
\figurenum{4}
\plottwo{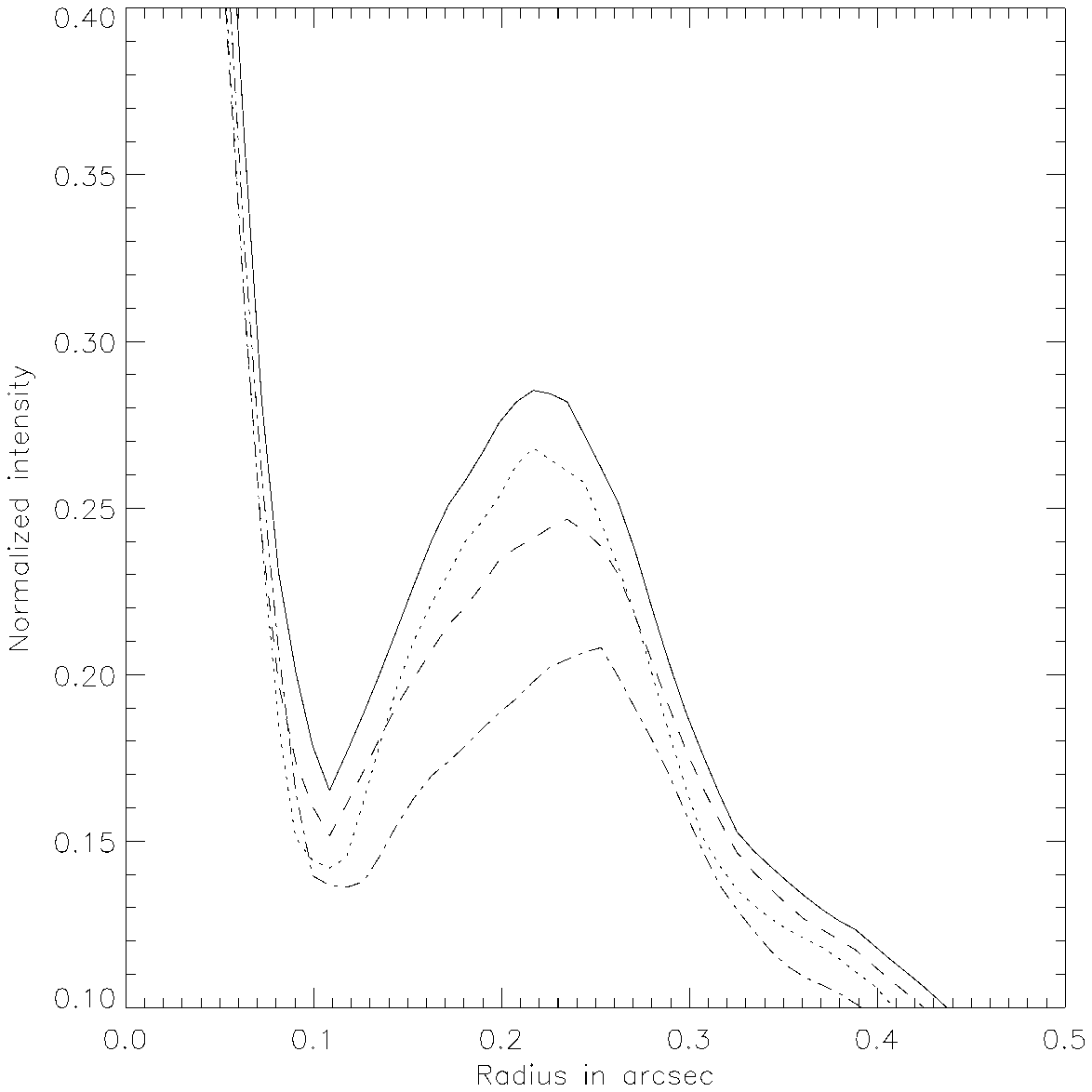}{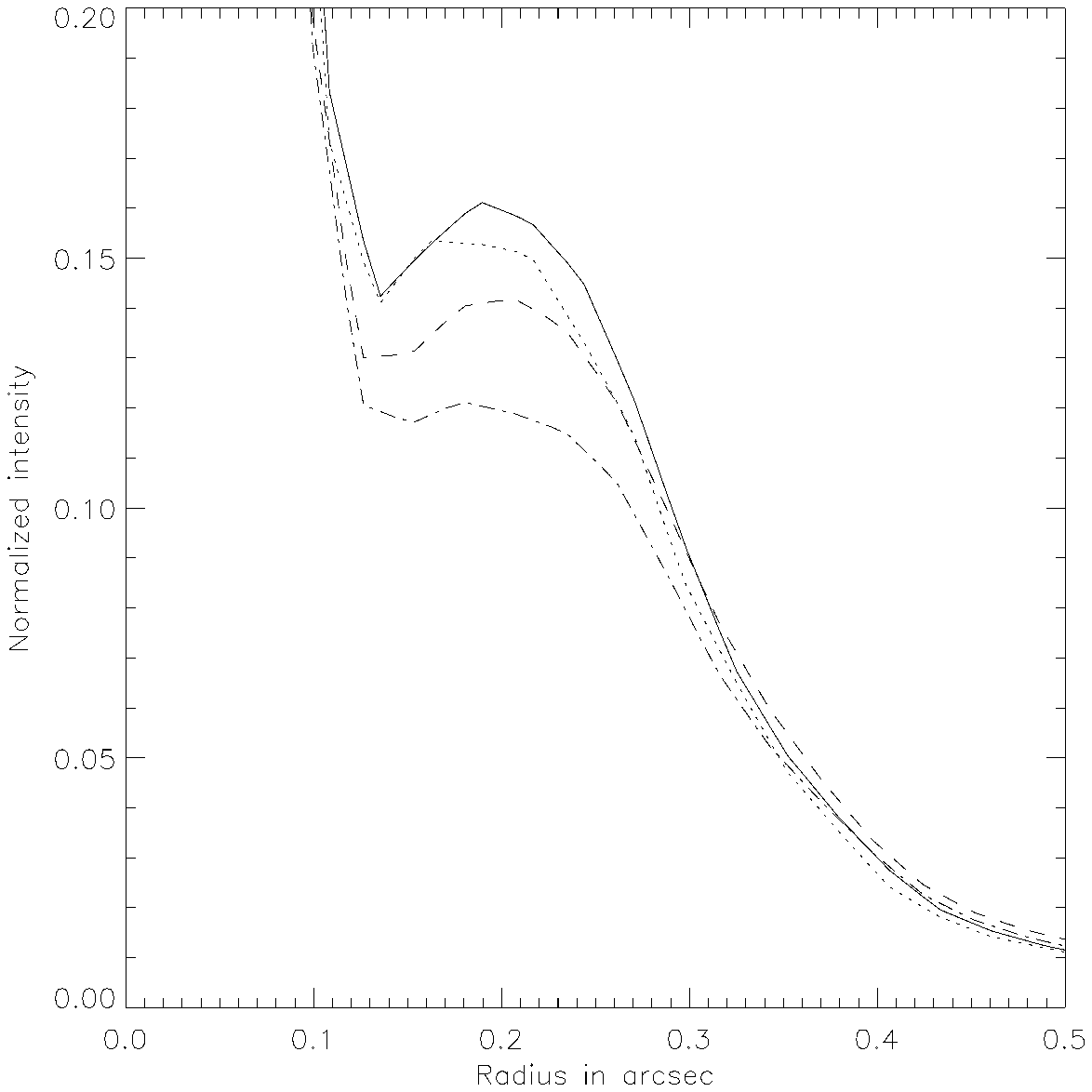}  
\caption{The normalized fluxes vs distance from the star for knot C$^{'}$ (left panel) and knot D$^{'}$ (right panel) measured at different times: in order, from the upper solid curve, 2002.88, 2003.95, 2004.99 to 2005.98, the lower dash-dot-dash line.}
\end{figure}

\begin{figure}
\figurenum{5}
\plotone{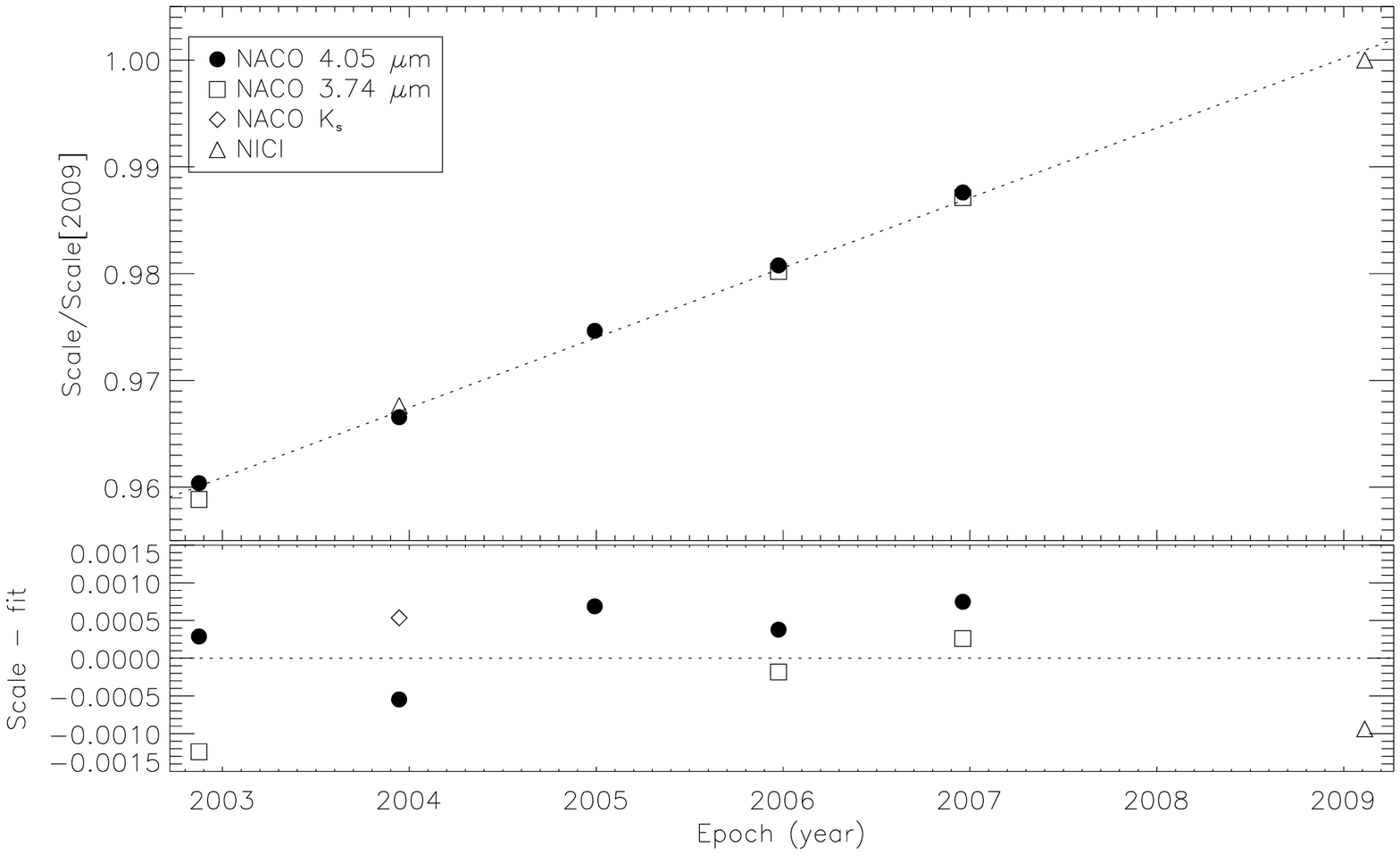}
\caption{The scaling factor between the NaCO and NICI images as a function of time. Using the 
measured scaling factors with the known pixel scales, we find that the Homunculus lobes were ejected in 1850$\pm$4.0.}
\end{figure}

\begin{figure}
\figurenum{6}
\plotone{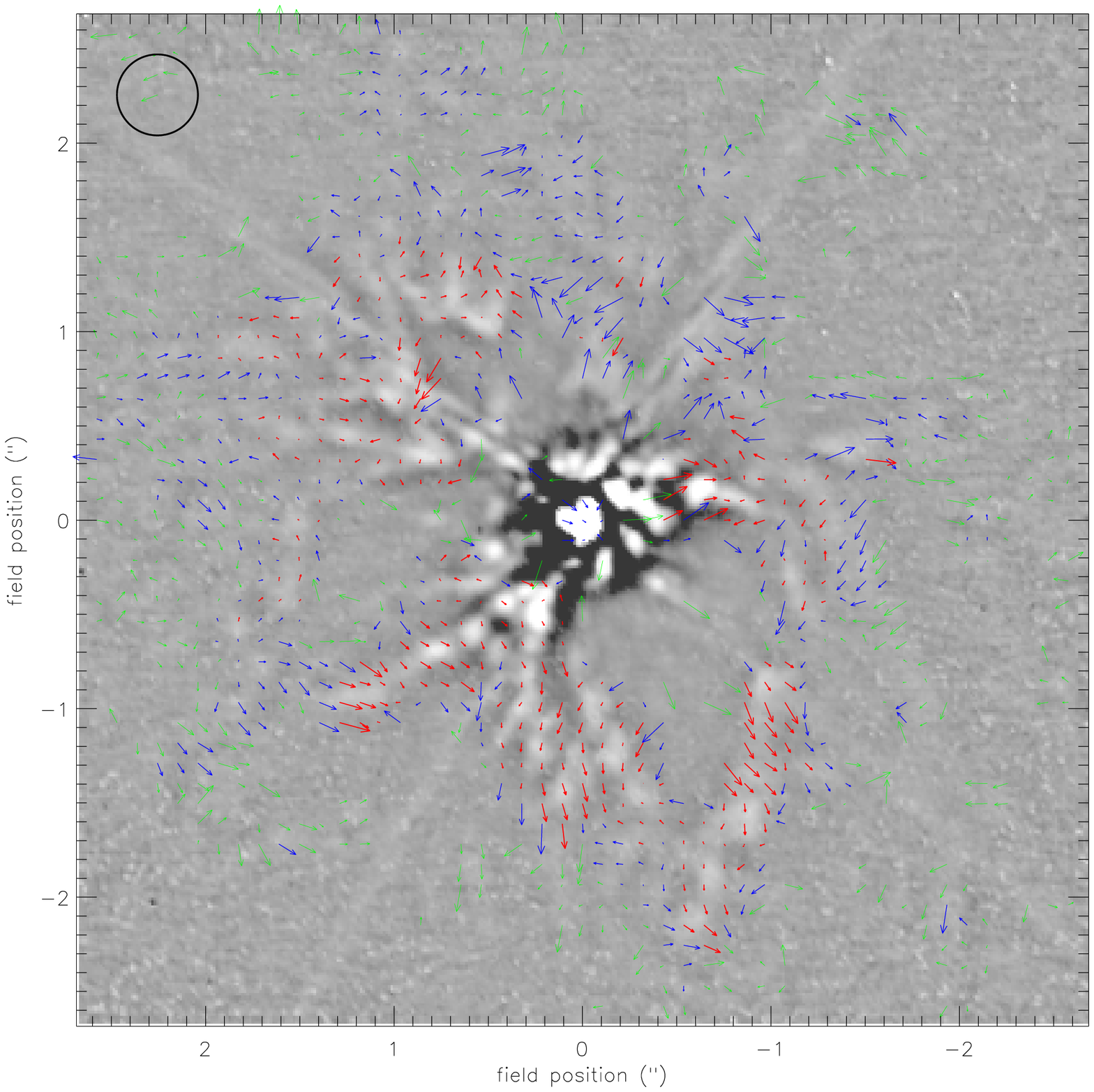}
\caption{Plot of the relative, co-moving  vector motions between the NICI (2009.1) and the NaCO (2002.9)  Ks-continuum images. The length of the arrows is proportional to the size of the
total motion. The arrows are color coded by the significance of the correlation (see text); 
red, 0.8 to 1.0; blue, 0.6 to 0.8 and green, 0.4 to 0.6. The circle in the upper left is the size of 
the correlation region. Also see Figure 8.}
\end{figure}

\newpage 

\begin{figure}
\figurenum{7}
\plotone{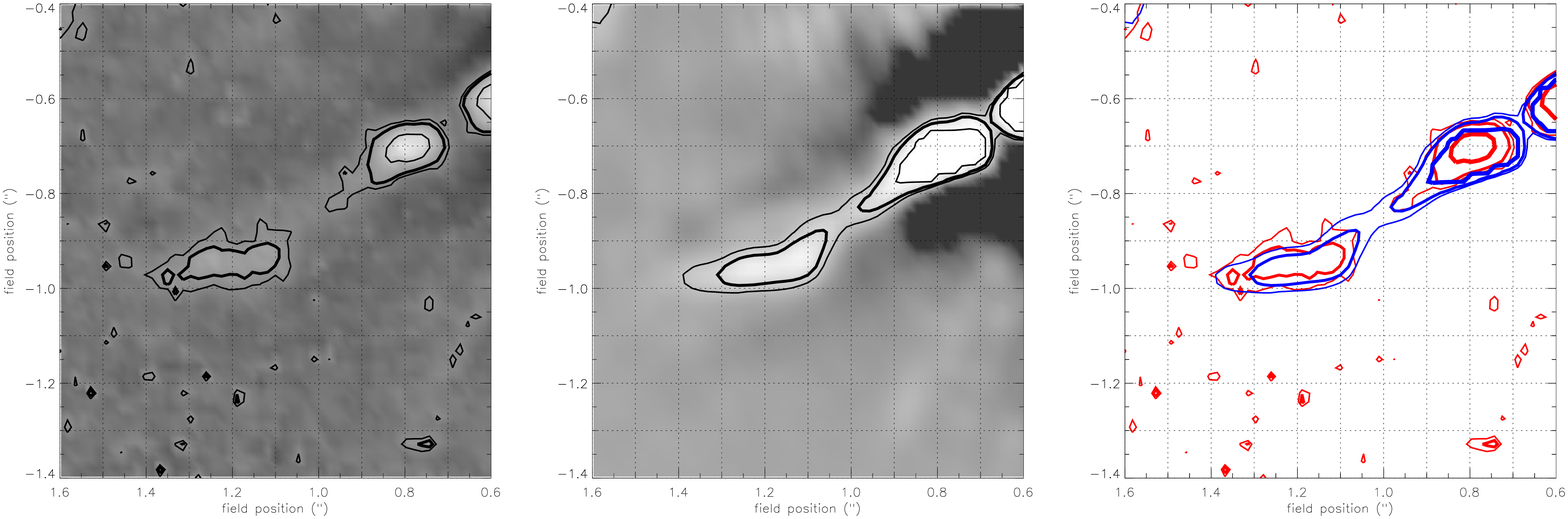}
\caption{Example of a filamentary structure showing a significant  motion 
between NaCO Ks (left) and NICI (middle) images. The superposition of the 
NICI and NaCO contour plots (right) illustrates the motion of the filamentary 
structure. The South-Eastern tip (at x=1.2”, y =-1.0”) of the filament has 
a tranverse motion of about 8 mas/yr while the bright knot (at x=0.8”, y =-0.7”) has a smaller (about 4 mas/yr) tranverse motion in the same direction.}
\end{figure}

\begin{figure}
\figurenum{8}
\plotone{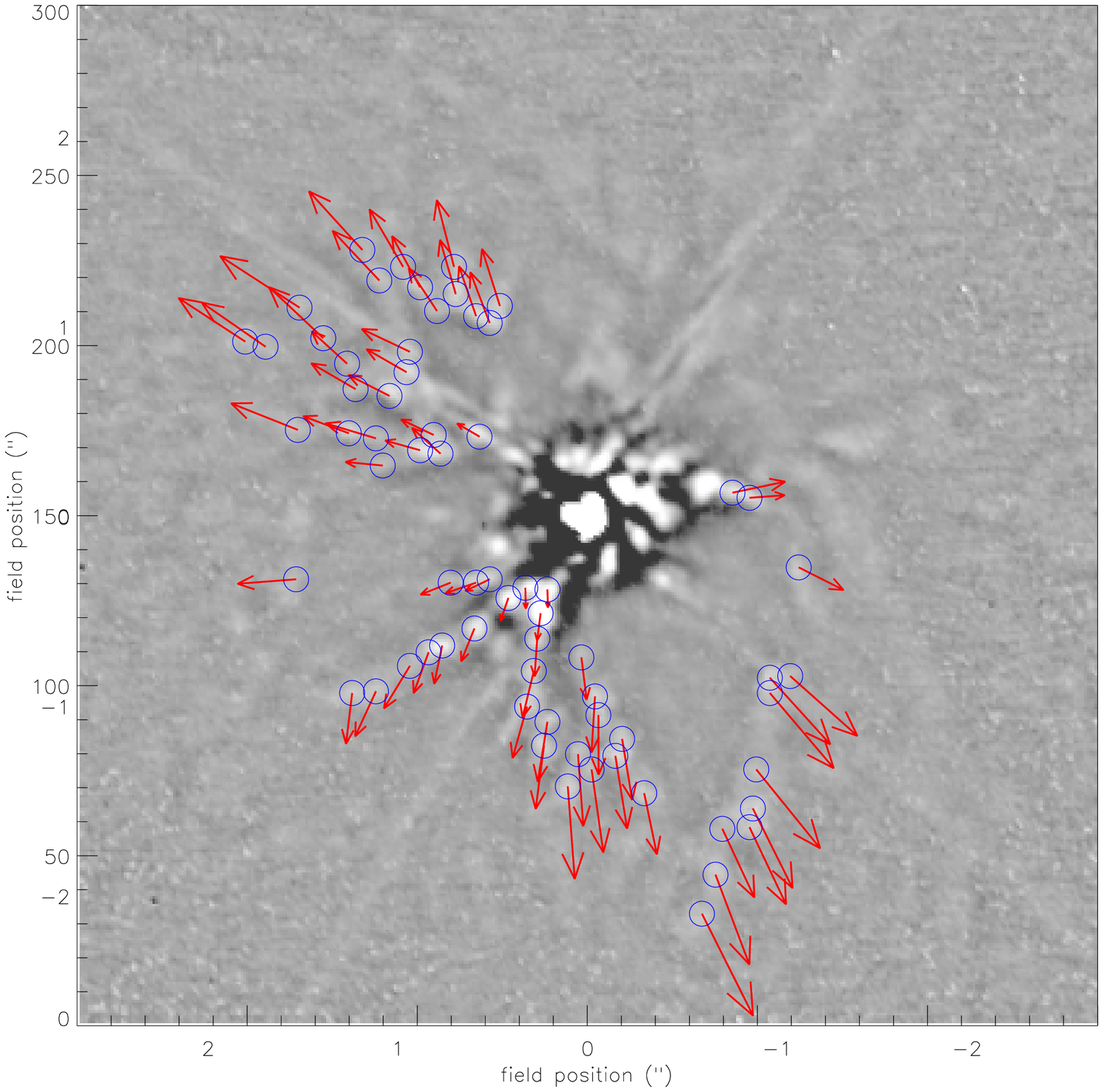}
\caption{A map of the apertures and the vectors for the true or intrinsic motions in the fixed 
{\it x,y} grid (see Appendix A) plotted on the NICI K-band continuum image. The arrow lengths  
are  proportional to the total motion.}
\end{figure}

\newpage

\begin{figure}
\figurenum{9}
\plotone{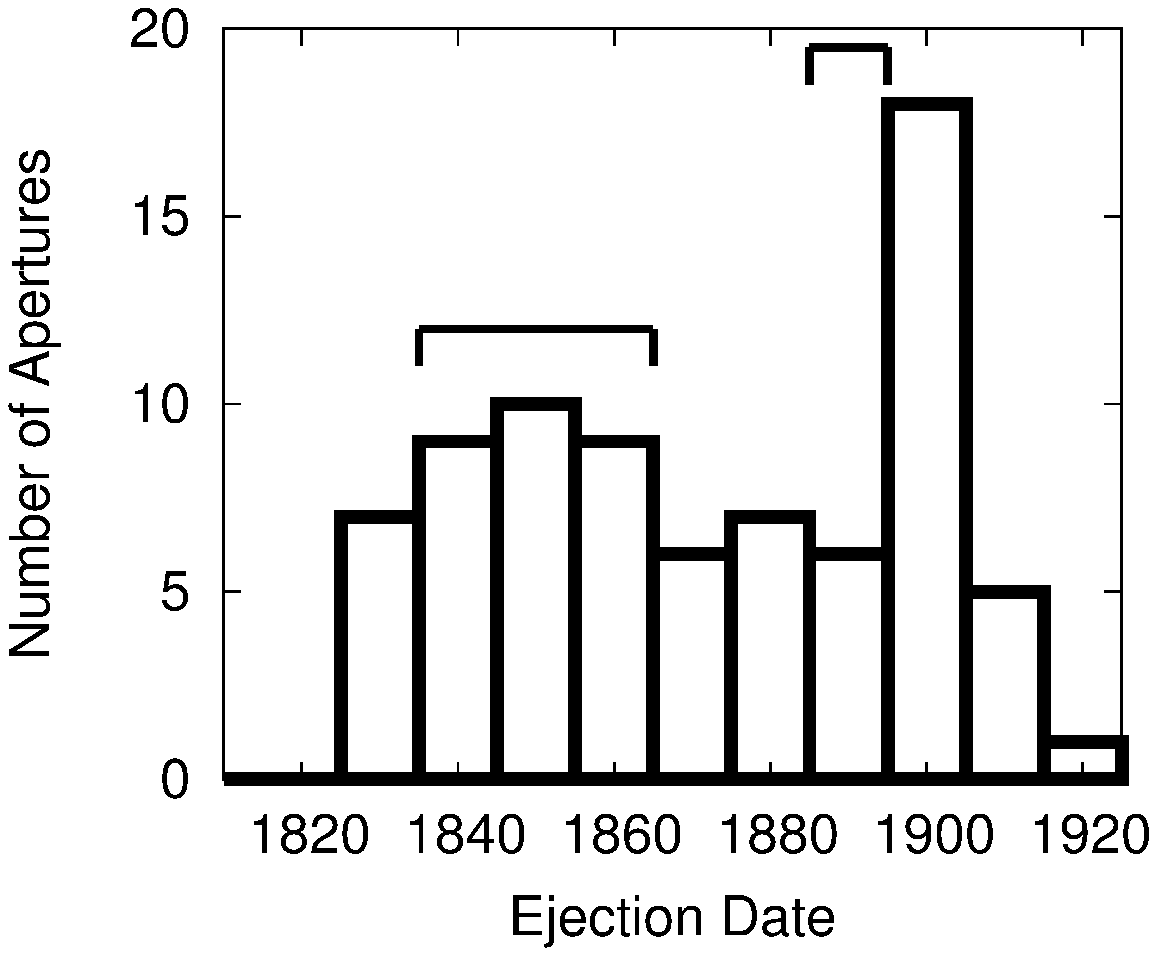}
\caption{The distribution of the ejection epochs from the  motions in Table 2.
The brackets indicate the times of the great eruption (1838 - 1858) and the second
eruption (1887 - 1895). The mean uncertainty in the ejection epochs is 7 years. Most of the apertures have an uncertainty of 5 to 10 years.   The errors are quoted in Table 2.} 
\end{figure} 

\newpage

\begin{figure}
\figurenum{10}
\plotone{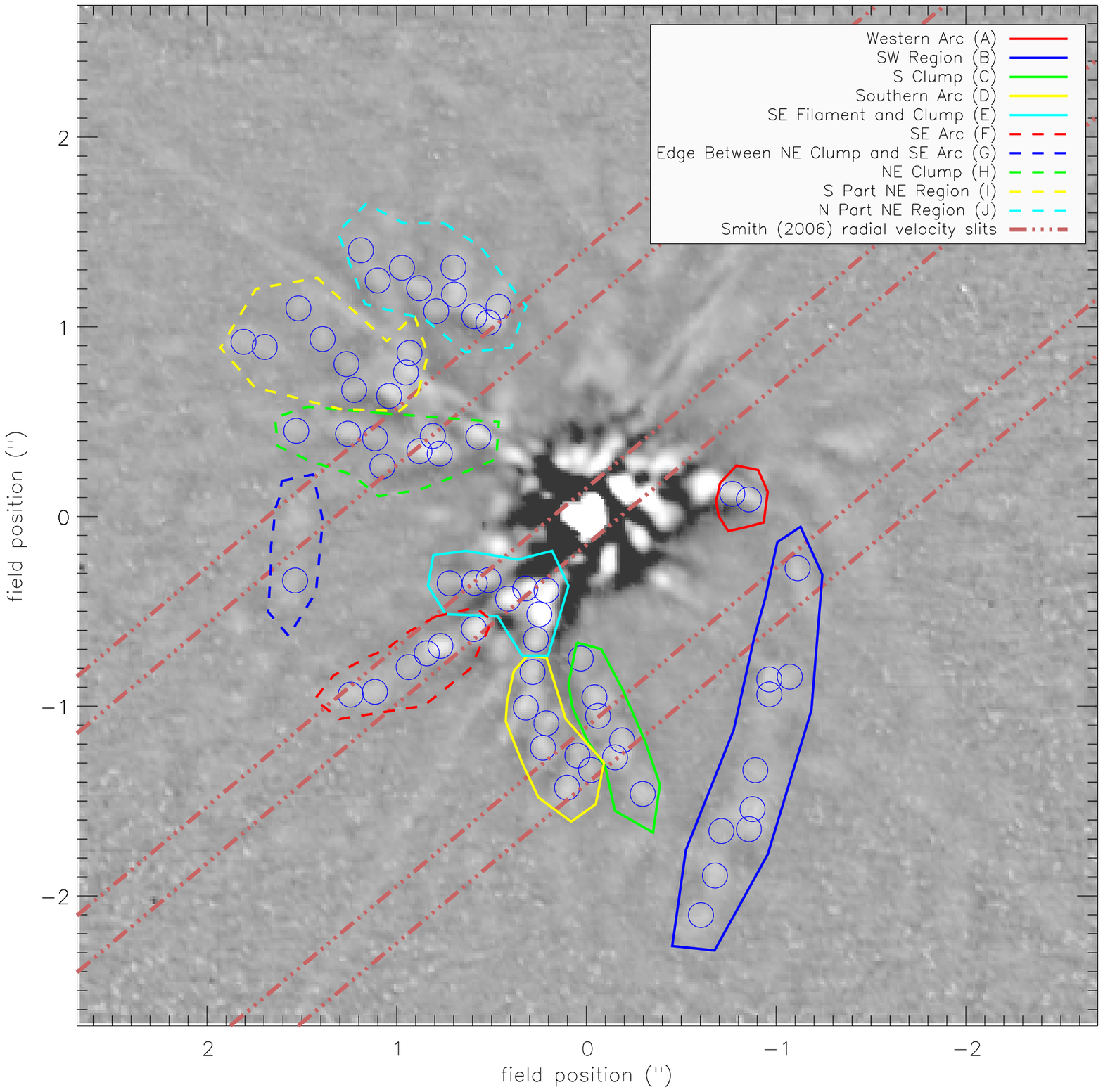}
\caption{A map of the larger regions and apertures (see Table 2) plotted on the NICI K-band continuum image. The parallel dashed lines from lower left to upper right note the locations of the slits used to measure radial velocities from Smith (2006).}
\end{figure}

\begin{figure}
\figurenum{11}
\plotone{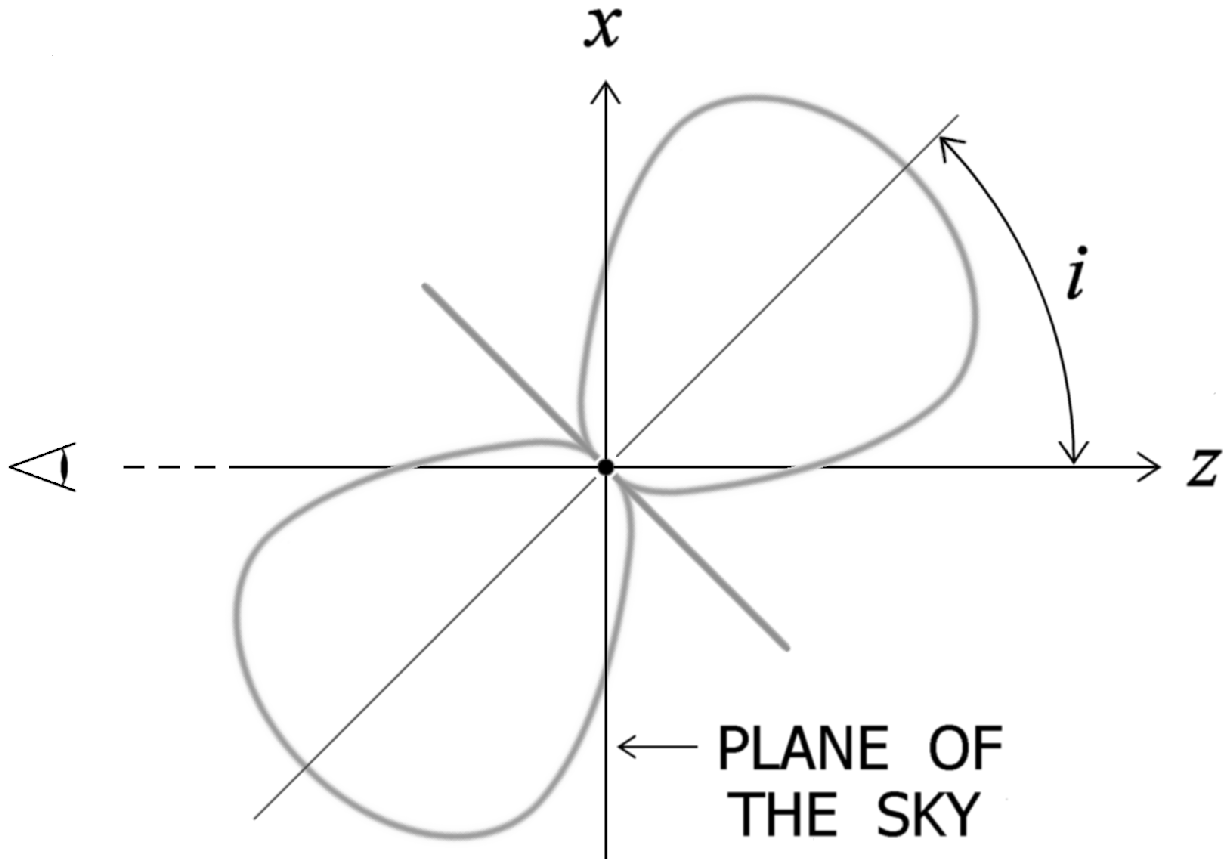}
\caption{The inclination $i$  angle ($40.7\arcdeg$) between our line of sight 
   and the axis of the bipolar Homunculus, adapted from Davidson et al. (2001).  
   This agrees with the conventional definition of $i$ for binary star orbits, 
   if we substitute the Homunculus equatorial plane for the orbit plane.  
   The total polar diameter of the Homunculus is roughly 40000 AU, 
  much larger than the area shown in Fig.\ 11. Caveat:  In fact the 
   axial symmetry is only an approximation.} 
\end{figure}

\begin{figure}
\figurenum{12}
\epsscale{0.8} 
\plotone{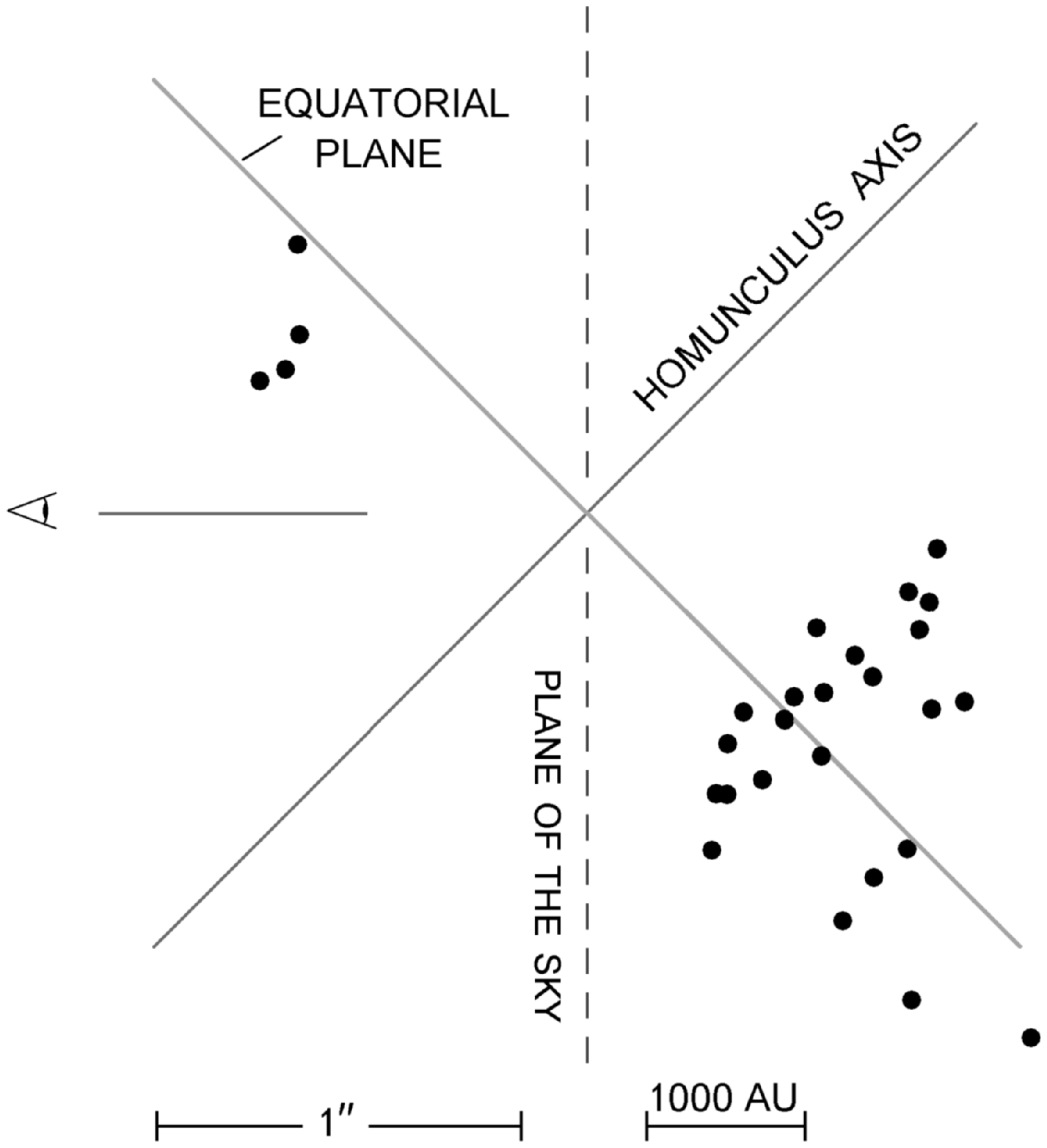}
\caption{``Side view'' of the aperture positions listed in Table 2 and shown in Fig.\ 8.  
These are 3-D locations projected into the $xz$ plane defined in Fig.\ 11, but the 
area shown here is much smaller.  Since this is a projected view rather than a 
 cross-section, random samples of a circular or toroidal configuration  
 would create a band connecting the upper left and lower right 
sets of points;  so the distribution plotted here does not provide 
 evidence for such a structure.}
\end{figure}



\begin{thebibliography}{}

\bibitem[Biller et al.(2004)]{Bill04} Biller, B.~A., Close, 
L., Lenzen, R., Brandner, W., McCarthy, D.~W., Nielsen, E., 
\& Hartung, M.\ 2004, \procspie, 5490, 389

\bibitem[Chesneau et al.(2005)]{Ches05} Chesneau, O., et al.\ 2005, \aap, 435, 1043

\bibitem[Chun et al.(2008)]{Chun08} Chun, M., et al.\ 2008, 
\procspie, 7015, 49

\bibitem[Currie et al.(1996)]{Currie96} Currie, D.~G., Dowling, 
D.~M., Shaya, E.~J., Hester, J.~J., The HST WF/Pc Instrument Definition 
Team, 
\& The HST WFPC2 Instrument Definition Team 1996, The Role of Dust in the Format
ion of Stars, 89

\bibitem[Currie 
\& Dowling(1999)]{Currie99} Currie, D.~G., \& Dowling, D.~M.\ 1999, in ASP Conf. Ser 179, ``Eta Carinae at the Millenium'', ed. J. Morse,
R. Humphreys \& A. Damineli (San Francisco, ASP), p. 72  

\bibitem[Davidson et al.(1995)]{KD95} Davidson, K., Ebbets, 
D., Weigelt, G., Humphreys, R.~M., Hajian, A.~R., Walborn, N.~R., 
\& Rosa, M.\ 1995, \aj, 109, 1784 

\bibitem[Davidson et al.(1997)]{KD97} Davidson, K., Ebbets, 
D., Johansson, S., Morse, J.~A., \& Hamann, F.~W.\ 1997, \aj, 113, 335 

\bibitem[Davidson \& Humphreys(1997)]{DH97} Davidson, K., \& Humphreys, R.~M.\ 1997, \araa, 35, 1 

\bibitem[Davidson et al.(2001)]{KD01} Davidson, K., Smith, N., Gull, T.~R., Ishibashi, K., \& Hillier, D.~J.\ 2001, \aj, 121, 1569 

\bibitem[Dorland et al.(2004)]{Dor04} Dorland, B.~N., Currie, 
D.~G., \& Hajian, A.~R.\ 2004, \aj, 127, 1052

\bibitem[Duschl et al.(1995)]{Duschl} Duschl, W.~J., Hofmann, 
K.-H., Rigaut, F., \& Weigelt, G.\ 1995, Revista Mexicana de Astronomia y Astrofisica Conference Se
ries, 2, 17 

\bibitem[Falcke et al.(1996)]{Weig96} Falcke, H., Davidson, K., Hofmann, K.-H., \& Weigelt, G.\ 199
6, \aap, 306, L17

\bibitem[Ftaclas et al.(2003)]{Ftac03} Ftaclas, C., 
Mart{\'{\i}}n, E.~L., \& Toomey, D.\ 2003, Brown Dwarfs, 211, 521

\bibitem[Gaviola(1950)]{Gav50} Gaviola, E.\ 1950, \apj, 111, 408 

\bibitem[Humphreys et al.(1999)]{RMH99} Humphreys, R.~M., 
Davidson, K., \& Smith, N.\ 1999, \pasp, 111, 1124 

\bibitem[Ishibashi et al.(2003)]{Bish03} Ishibashi, K., et 
al.\ 2003, \aj, 125, 3222 

\bibitem[Ishibashi(2005)]{Bish05} Ishibashi, K.\ 2005, The 
Fate of the Most Massive Stars, 332, 131

\bibitem[Lenzen et al.(2004)]{Lenz04} Lenzen, R., Close, L., 
Brandner, W., Biller, B., \& Hartung, M.\ 2004, \procspie, 5492, 970 

\bibitem[Marois et al.(2005)]{Mar05} Marois, C., Doyon, R., 
Nadeau, D., Racine, R., Riopel, M., Vall{\'e}e, P., 
\& Lafreni{\`e}re, D.\ 2005, \pasp, 117, 745 

\bibitem[Marois et al.(2006)]{Mar06} Marois, C., 
Lafreni{\`e}re, D., Macintosh, B., \& Doyon, R.\ 2006, \apj, 647, 612

\bibitem[Martin et al.(2006)]{JCM} Martin, J.~C., Davidson, 
K., \& Koppelman, M.~D.\ 2006, \aj, 132, 2717 

\bibitem[Meaburn et al.(1987)]{Meaburn87} Meaburn, J., Wolstencroft, R.~D., \& W
alsh, J.~R.\ 1987, \aap, 181, 333

\bibitem[Meaburn et al.(1993)]{Meaburn93} Meaburn, J., Gehring, G., Walsh, J.~R.
, Palmer, J.~W., Lopez, J.~A., Bryce, M., \& Raga, A.~C.\ 1993, \aap, 276, L21 

\bibitem[Meaburn et al.(1996)]{Meaburn96} Meaburn, J., Boumis, 
P., Walsh, J.~R., Steffen, W., Holloway, A.~J., Williams, R.~J.~R., 
\& Bryce, M.\ 1996, \mnras, 282, 1313 

\bibitem[Mehner et al.(2010)]{Mehner}Mehner, A., Davidson, K., Ferland, G. J. \& Humphreys, R. M., \ 2010, \apj, 710, 729 

\bibitem[Morse et al.(1998)]{Morse1998} Morse, J.~A., Davidson, 
K., Bally, J., Ebbets, D., Balick, B., \& Frank, A.\ 1998, \aj, 116, 2443

\bibitem[Morse et al.(2001)]{Morse01} Morse, J.~A., Kellogg, 
J.~R., Bally, J., Davidson, K., Balick, B., 
\& Ebbets, D.\ 2001, \apjl, 548, L207 

\bibitem[Polomski et al.(1999)]{Pol99} Polomski, E.~F., 
Telesco, C.~M., Pi{\~n}a, R.~K., \& Fisher, R.~S.\ 1999, \aj, 118, 2369 

\bibitem[Rigaut \& Gehring(1995)]{RG95} Rigaut, F., \& Gehring, G.\ 1995, Revist
a Mexicana de Astronomia y Astrofisica Conference Series, 2, 27 

\bibitem[Ringuelet(1958)]{Ring58} Ringuelet, A.~E.\ 1958, \zap, 46, 276 

\bibitem[Smith et al.(1995)]{CSmith95}Smith, C. H., Aitken, D. K., Moore, T. J. T., 
Roche, P. F., Pueter, R. C., \& Pina, R. K. \ 1995, \mnras, 273, 354

\bibitem[Smith \& Gehrz(1998)]{Smith98} Smith, N., \& Gehrz, R.~D.\ 1998, \aj, 1
16, 823

\bibitem[Smith et al.(1998)]{SGK98} Smith, N., Gehrz, R.~D., 
\& Krautter, J.\ 1998, \aj, 116, 1332

\bibitem[Smith \& Gehrz(2000)]{Smith00} Smith, N., \& Gehrz, R.~D.\ 2000, \apjl,
 529, L99 

\bibitem[Smith \& Davidson(2001)]{Smith01} Smith, N., \& Davidson, K.\ 2001, \apjl, 551, L101

\bibitem[Smith et al.(2002)]{Smith02b} Smith, N., Gehrz, R.~D., 
 Hinz, P.~M., Hoffmann, W.~F., Mamajek, E.~E., Meyer, M.~R., 
 \& Hora, J.~L.\ 2002, \apjl, 567, L77

\bibitem[Smith(2002)]{Smith02} Smith, N.\ 2002, \mnras, 337, 1252

\bibitem[Smith et al.(2003)]{Smith03} Smith, N., Gehrz, R.~D., 
 Hinz, P.~M., Hoffmann, W.~F., Hora, J.~L., Mamajek, E.~E., 
 \& Meyer, M.~R.\ 2003, \aj, 125, 1458 

\bibitem[Smith et al.(2004)]{Smith04} Smith, N., et al.\ 2004, \apj, 605, 405 

\bibitem[Smith(2005)]{Smith05} Smith, N.\ 2005, \mnras, 357, 1330

\bibitem[Smith(2006)]{Smith06} Smith, N.\ 2006, \apj, 644, 1151

\bibitem[Teodoro et al.(2008)]{Teodoro} Teodoro, M., Damineli, 
 A., Sharp, R.~G., Groh, J.~H., \& Barbosa, C.~L.\ 2008, \mnras, 387, 564

\bibitem[Weigelt \& Ebersberger(1986)]{Weig86} Weigelt, G., \& Ebersberger, J.\ 
 1986, \aap, 163, L5

\bibitem[Weigelt et al.(1995)]{Weig95} Weigelt, G., et al.\ 
 1995, Revista Mexicana de Astronomia y Astrofisica Conference Series, 2, 11

\bibitem[Zethson et al.(1999)]{Zeth99} Zethson, T., Johansson, S., Davidson, K.,
  Humphreys, R.~M., Ishibashi, K., \& Ebbets, D.\ 1999, \aap, 344, 211


\end{thebibliography}
\end{document}